\documentclass{emulateapj}

\def\myputfigure#1#2#3#4#5%
{\vskip#5pt\makebox[0pt]{\hskip#2in
\includegraphics[width=#3\textwidth]{#1}}\vskip#4pt\hfill}

\newcommand\lsim{\mathrel{\rlap{\lower4pt\hbox{\hskip1pt$\sim$}}
        \raise1pt\hbox{$<$}}}
\newcommand\gsim{\mathrel{\rlap{\lower4pt\hbox{\hskip1pt$\sim$}}
        \raise1pt\hbox{$>$}}}

\newcommand{\nf}{x_{\rm HI}}
\newcommand{\nfIGM}{x_{\rm HI}^{\rm IGM}}

\newcommand{\strom}{HII region}
\newcommand{\lya}{Lyman~$\alpha$}

\newcommand{\taudamp}{\tau_{D}}
\newcommand{\taures}{\tau_{R}}
\newcommand{\lobs}{\lambda_{\rm obs}}
\newcommand{\tauobs}{\tau_{\rm obs}}
\newcommand{\tausim}{\tau_{\rm sim}}
\newcommand{\zsource}{z_{\rm Q}}

\newcommand{\qnamesfourtwo}{SDSS J1148+5251}
\newcommand{\qnamestwoeight}{SDSS J1030+0524}
\newcommand{\qnamestwotwo}{SDSS J1623+3112}

\newcommand{\qnamefourtwo}{J1148+5251}
\newcommand{\qnametwoeight}{J1030+0524}
\newcommand{\qnametwotwo}{J1623+3112}
\newcommand{\qnametwo}{J1048+4637}

\newcommand{\tautot}{\tau_{\rm Ly\alpha}}

\newcommand{\fgamma}{f_\Gamma}
\newcommand{\volavenf}{\bar{x}_{\rm HI}}

\begin{document}

\submitted{ApJ, in press}
\title{Constraints on Reionization and Source Properties from the Absorption Spectra of $z>6.2$ Quasars}

\author{Andrei Mesinger\altaffilmark{1, 2} \& Zolt\'{a}n Haiman\altaffilmark{2}}

\altaffiltext{1}{Yale Center for Astronomy and Astrophysics, Yale University, 260 Whitney Avenue, New Haven, CT 06520}
\altaffiltext{2}{Department of Astronomy, Columbia University, 550 West 120th Street, New York, NY 10027}

\begin{abstract}
We make use of hydrodynamical simulations of the intergalactic medium
(IGM) to create model quasar absorption spectra.  We compare these
model spectra with the observed Keck spectra of three $z>6.2$ quasars
with full Gunn-Peterson troughs: \qnamesfourtwo\ ($z=6.42$),
\qnamestwoeight\ ($z=6.28$), and \qnamestwotwo\ ($z=6.22$). We fit the
probability density distributions (PDFs) of the observed \lya\ optical
depths ($\tau_\alpha$) with those generated from the simulation, by
adopting a template for the quasar's intrinsic spectral shape, and
exploring a range of values for the size of the quasar's surrounding
HII region, $R_S$, the volume-weighted mean neutral hydrogen fraction
in the ambient IGM, $\volavenf$, and the quasar's ionizing photon
emissivity, $\dot N_Q$.  In order to avoid averaging over possibly
large sightline--to--sightline fluctuations in IGM properties, we
analyze each observed quasar independently.
We find the following results for \qnamefourtwo, \qnametwoeight, and
\qnametwotwo:
The best--fit sizes $R_S$ are 40, 41, and 29 (comoving) Mpc,
respectively.  For the later two quasars, the value is significantly
larger than the radius corresponding to the wavelength at which the
quasar's flux vanishes. These constraints are tight, with only $\sim
10\%$ uncertainties, comparable to those caused by
redshift--determination errors.
The best--fit values of $\dot N_Q$ are 2.1, 1.3, and 0.9 $\times$
10$^{57}$ s$^{-1}$, respectively, with a factor of $\sim$ 2
uncertainty in each case.  These values are a factor of $\sim$ 5
lower than expected from the template spectrum of Telfer et
al. (2002).
Finally, the best--fit values of $\volavenf$ are 0.16, 1.0, and 1.0,
respectively. The uncertainty in the case of \qnamefourtwo\ is large,
and $\volavenf$ is not well constrained. However, for both
\qnametwoeight\ and \qnametwotwo\, we find a significant lower limit
of $\volavenf\gsim 0.033$.  
Our method is different from previous analyses of the GP absorption
spectra of these quasars, and our results strengthen the evidence that
the rapid end--stage of reionization is occurring near $z\sim 6$.
\end{abstract}


\section{Introduction}
\label{sec:intro}

The epoch of reionization, when the radiation from early
generations of astrophysical objects ionized the intergalactic medium
(IGM), offers a wealth of information about cosmological structure
formation and physical processes in the early universe.  Only
recently have we begun to gather clues concerning this epoch.  The
Thomson scattering optical depth, $\tau_e = 0.09\pm0.03$
\citep{Page06, Spergel06}, measured from the polarization anisotropies
in the cosmic microwave background (CMB) by the {\it Wilkenson
Microwave Anisotropy Probe} ({\it WMAP}), suggests that cosmic
reionization began at redshift $z\gsim10$.  However, the current
uncertainty of this value, and the fact that the optical depth
measures only an integrated electron column density, means that many
possible competing reionization scenarios can still be accommodated.

In order to chart the progress of reionization, several attempts have
been made to measure the neutral fraction of the IGM at high redshifts
($z\gsim6$).  The detection of dark spectral regions, so-called
Gunn-Peterson (GP) troughs, in the Lyman line absorption spectra of
high-redshift quasars discovered in the Sloan Digital Sky Survey
(SDSS) offers a probe of the late stages of reionization.  The
Ly$\alpha$ optical depths associated with such GP troughs, and more
importantly their accelerated evolution at $z\gsim6$, suggests that
the reionization epoch is ending at $z\sim6$, with a lower
limit on the volume-weighted IGM neutral fraction, $\volavenf(z\sim6)
\gsim 10^{-3}$, along at least the line of sight (LOS) to two quasars
\citep{White03, Fan06}.  The sharp decline in flux at the edge of the
GP trough of quasar \qnametwoeight\ can be used to infer the stronger
constraint of $\volavenf(z\sim6) \gsim 0.2$, along that LOS \citep{MH04}.  The observed size of the quasars' surrounding HII
regions can be used to infer a similar, independent lower limit
\citep{WL04_nf, MH04}, although this requires additional assumptions
concerning the quasar and its environment, and the inferred neutral
fraction scales directly with these assumptions \citep{BH06,
Maselli06}.  The rapid evolution in the sizes of the HII regions
for different quasars between $z\sim5.7$ and $z\sim6.4$ suggests that
the average neutral fraction increased by a factor of $\sim 10$ during
this interval \citep{Fan06}, although there are alternative ways to
account for the steep size--evolution \citep{BH06}.  The distribution
of spectral sizes of dark gaps should help distinguish these scenarios
in the future \citep{GCF06, Fan06}.  Upper limits on the neutral
fraction at $z\sim6$ have also been suggested by other means.  The
lack of significant evolution in the luminosity function (LF) of \lya\
emitters between $z\approx5.7$ and $z\approx6.5$ was used to infer
$\volavenf(z\sim6) \lsim 0.3$ \citep{MR04, HC05}, if the galaxies are
assumed to lie in isolation, or $\volavenf(z\sim6) \lsim 0.5$
\citep{FZH06}, if an analytic prescription of the clustering of HII
bubbles \citep{FZH04} is taken into account. However, a recent, more
accurate measurement of the LF at $z\approx 6.5$ revealed evidence of
a decrease, weakening these upper limits \citep{Kashikawa06}.
Finally, the lack of noticeable absorption due to the Ly$\alpha$
damping wing in the afterglow spectrum of the gamma-ray burst 050904
at $z = 6.3$ can constrain $\volavenf(z\sim6) \lsim 0.2$ along that
LOS \citep{Totani06}.

The preponderance of constraints mentioned above suggests that the
combination of all existing data is consistent with $\volavenf(z\sim6)
\sim 0.1$. While this is plausible, the constraints listed above all
rely on various model assumptions.  For example, the conversion of an
observed average flux decrement to a neutral fraction \citep{Fan02,
White03, Fan06} is difficult, and requires precise ab--initio
knowledge of the IGM density distribution at high-redshifts
\citep{OF05, Fan06}.  All of the other constraints mentioned above
rely on the IGM density distribution, as well, albeit less
sensitively.  Techniques involving the sizes of HII regions
\citep{WL04_nf, Maselli06, Fan06} are subject to degeneracies
between quasar lifetime, ionizing flux, and the IGM neutral fraction.
Further complicating matters for the HII bubble size techniques is the
fact that the onset of the GP trough need not correspond to the edge
of the HII region.  The GP through just corresponds to the region
where the flux drops below the detection threshold, and thus merely
provides a {\it lower limit} on the size of the HII region
(\citealt{MH04, Maselli06}; see also \citealt{BH06} for a recent,
comprehensive review of these issues).  Furthermore, determining the
number density and evolution of Ly$\alpha$ emitting galaxies is
complicated and depends on knowledge of galaxy clustering \citep{HC05, FHZ04,
FZH06}, absorber clustering, as well as the large-scale density
(e.g. \citealt{BL04, BL04_grbvsqso}) and velocity fields
(e.g. \citealt{CHM05}), which are all unconstrained at the relevant
high redshifts.

More generally, the drawback of many of these methods is that they
rely on the combination of data from multiple LOSs to obtain
statistical significance.  However, assuming constancy in the IGM
properties along disparate LOSs is dangerous at high-redshifts,
especially approaching the end of reionization, when large
sightline-to-sightline fluctuations are expected in the density field
and the ionizing background (see, e.g., recent reviews by
\citealt{DBM06} and \citealt{LOF06}).  Indeed, the current sample of high-redshift
quasars already reveals large fluctuations in the IGM properties along
different LOSs \citep{Fan06}.  This would suggest that the most
rewarding technique would be able to analyze and obtain constraints
from each source independently from other sources.  This is
statistically less powerful (e.g. \citealt{MHC04}), but the rewards
can be rich, potentially providing additional constraints on the
properties of individual quasars and their environments, such as the
quasar's emissivity of ionizing photons, size of its surrounding HII
region, sharpness of the HII region boundary, the quasars'
X-ray emission, as well as a measure of the sightline-to-sightline fluctuations in these quantities \citep{MH04}.

{\it The purpose of this paper is to derive these parameters, i.e.
the hydrogen neutral fraction, the size of surrounding HII region, and
the ionizing emissivity, by separately modeling the spectra of each of
the $z>6.2$ quasars.}  Specifically, we
analyze \qnamefourtwo\ ($z=6.42$), \qnametwoeight\ ($z=6.28$) and
\qnametwotwo\ ($z=6.22$).  At present, a 4th quasar is known with a
full GP trough.  This source, however, is a broad absorption line
(BAL) quasar \qnametwo\ ($z=6.2$), with complex spectral features
(Fan, X., private communication) that preclude a simple interpretation
in terms of IGM properties. We have therefore omitted this quasar from
our study.

Our analysis differs from our previous work \citep{MH04} where we
modeled the gross transmission features in Ly$\alpha$ and Ly$\beta$
close to the edge of the HII region surrounding \qnametwoeight.  In
the present study, we perform a much more detailed analysis, by
fitting the quasar's intrinsic emission, and modeling the distribution
of optical depths, pixel--by--pixel, blueward of the Ly$\alpha$ line
center, following the procedure proposed in \citet{MHC04}.  Unlike in
\citet{MH04}, here we do not make use of the Ly$\beta$ absorption
spectrum, except to set a lower limit in the allowed size of the HII
region surrounding \qnametwoeight, the only one among the three
quasars that has a noticeable offset between the Ly$\alpha$ and
Ly$\beta$ GP troughs.  Our reason for ignoring the Ly$\beta$ region is
that the majority of the pixels we analyze (see Figure \ref{fig:fits}
below) have flux detected in the Ly$\alpha$ absorption region, which
can directly be converted to a Ly$\alpha$ optical depth. This avoids
the need for a statistical modeling of the foreground Ly$\alpha$
absorption, necessary to convert Ly$\beta$ to a Ly$\alpha$ optical
depth.

This paper is organized as follows.  
In \S~\ref{sec:anal}, we describe our analysis technique, including
generating the simulated absorption (\S~\ref{sec:sim}), fitting for
the quasars' intrinsic emission spectrum (\S~\ref{sec:emission}), and
the statistical method used to compare the observed and simulated
spectra (\S~\ref{sec:comp}).
In \S~\ref{sec:results}, we present our results for each of the
quasars we analyze.  
In \S~\ref{sec:discussion}, we discuss various aspects of our results,
such as the origin of the constraints we obtain, and the robustness
of our results in the face of uncertainties in the quasar template
spectrum.
In \S~\ref{sec:conc}, we summarize our key findings and present our
conclusions.

\section{Analysis}
\label{sec:anal}

In this section, we describe the three components of our analysis: (i)
a hydrodynamical simulation to model the absorption
(\S~\ref{sec:sim}); (ii) modeling the quasars' intrinsic emission
spectrum (\S~\ref{sec:emission}); and (iii) the statistical comparison
of observed and simulated spectra (\S~\ref{sec:comp}).  Unless stated
otherwise, we adopt the background cosmological parameters
($\Omega_\Lambda$, $\Omega_{\rm M}$, $\Omega_b$, n, $\sigma_8$, $H_0$)
= (0.76, 0.24, 0.0407, 1, 0.76, 72 km s$^{-1}$ Mpc$^{-1}$), consistent
with the three--year results by the {\it WMAP} satellite
\citep{Spergel05}.  We use redshift, $z$, and the observed wavelength,
$\lobs$, interchangeably throughout this paper as a measure of the
distance along the line of sight.  
Unless stated otherwise, we quote all quantities in comoving units.

\subsection{Simulating Absorption in the IGM}
\label{sec:sim}

In order to simulate the absorption spectra, we make use of a
hydrodynamical simulation box in a $\Lambda$CDM universe at redshift
$z$ = 6.  The details of the simulation are described in
\citet{Cen03code}, and we only briefly discuss the relevant parameters
here.  The box is 11 $h^{-1}$ Mpc on each side, with each pixel being
about 25.5 $h^{-1}$ kpc.  This scale resolves the Jeans length in the
smooth, partially-ionized\footnote{The temperature of a neutral IGM
before any ionization would be several orders of magnitude lower than
in the partially-ionized case, and the Jeans length in the IGM would
be under-resolved by a factor of $\sim$6.}  IGM by more then a factor
of 10.  The box also has a slightly different set of cosmological
parameters: $(\Omega_\Lambda, \Omega_M, \Omega_b, n, \sigma_8,$ $H_0)$
= (0.71, 0.29, 0.047, 1, 0.85\footnote{The three--year {\it WMAP} results yielded a smaller value of $\sigma_8 = 0.76 \pm 0.05$ \citep{Spergel06}.  A smaller $\sigma_8$ would result in a narrower distribution of optical depths than those obtained from the box we use.  Although this does not exactly mimic the effect of increased damping wing absorption from a more neutral universe, which shifts the optical depth distribution to higher values, it is likely that if confirmed, a smaller $\sigma_8$ would weaken our lower limits on the IGM neutral fraction below.  Estimates for the sensitivity of our results on our choice of cosmological parameters is quantitatively difficult without running a suite of models.  We note however that analysis combining the three--year {\it WMAP} results with data from the Lyman--$\alpha$ forest suggests a higher value of $\sigma_8 = 0.86 \pm 0.03$ \citep{Lewis06}, which is consistent with our choice.}
, 70 km s$^{-1}$ Mpc$^{-1}$).

Density and velocity information was extracted from the simulation box
along randomly chosen lines of sight.  The step size along each LOS
was taken to be 5.1 $h^{-1}$ kpc, which resolves the \lya\ Doppler
width in the partially-ionized IGM by more than a factor of 40.  The
exact value of the step size was chosen somewhat arbitrarily, and does
not influence the results as long as it adequately resolves the
Doppler width.  At each step, the density and velocity values were
averaged for the neighboring pixels, and weighted by the distance to
the center of the pixels.  We extended each LOS by the common practice
of randomly choosing a LOS through the box, and stacking several
segments together \citep{Cen94, MHC04, KGH05, BH06}.  The r.m.s. mass fluctuation on the scale the box is $\sigma(z=6.3) \sim 0.15$.  This is a small enough value that neglecting wavelengths longer than the box size is justified in our analysis.  Note however, that this is not the case for modeling highly non-linear processes, such as reionization (e.g. \citealt{BL04}). 
    When using such
a LOS in simulating a given quasar spectrum, density values from the
simulation were scaled as $(1+z)^3$.

The total optical depth due to \lya\ absorption, $\tau$, between an
observer at z = 0 and a source at z = $\zsource$, at an observed
wavelength of $\lobs = \lambda_0 (1+\zsource)$, is given by:

\begin{equation}
\label{eq:total_tau}
\tau(\lobs) = \int_{0}^{\zsource} dz ~ c \frac{dt}{dz} ~ n(z) ~ \nf(z) ~ \sigma \left[ \frac{\lobs}{(1+z)(1-\frac{v(z)}{c})} \right]
\end{equation}

\noindent where $n(z)$, $v(z)$, $\nf(z)$ are the total hydrogen number
density, the component of the peculiar velocity along the LOS,
and the local hydrogen neutral fraction, respectively, at redshift
$z$.  The Ly$\alpha$ absorption cross section, to first order in
$v(z)/c$, is given by $\sigma[\lobs/(1+z)/(1-v(z)/c)]$.

\begin{figure}
\vspace{+0\baselineskip}
\myputfigure{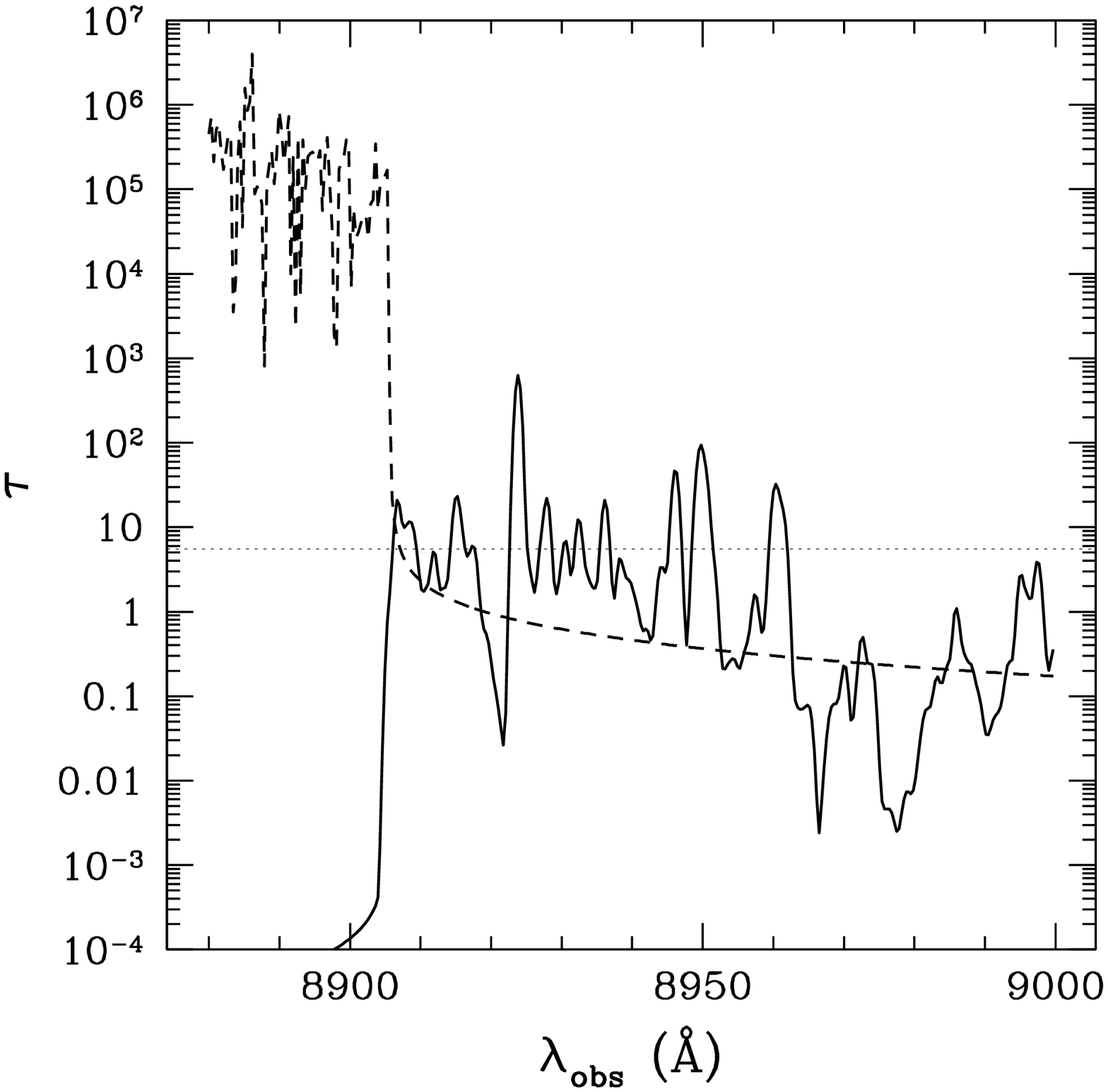}{3.3}{0.5}{.}{0.}  \figcaption{ Model
from a hydrodynamical simulation for the optical depth contributions
from within ($\tau_R$) and from outside ($\tau_D$) the local ionized
region for a typical line of sight towards a $\zsource=6.42$ quasar.
This figure was created with parameter choices ($R_S$, $\nfIGM$,
$\fgamma$) = (40 Mpc, 1, 1).  The dashed curve corresponds to
$\taudamp$, and the solid curve corresponds to $\taures$.  The total
\lya\ optical depth is the sum of these two contributions, $\tau$ =
$\taures$ + $\taudamp$.  The dotted line demarcates roughly the
maximum total optical depth, $\tau \sim 5.5$, at which a signal can be
detected with a confidence greater than 1-$\sigma$, in the spectral
region just blueward of the Ly$\alpha$ emission line (see discussion
in \S~\ref{sec:comp}).  In our analysis, the transmission
($e^{-\tau}$) is averaged over $\sim$ 3 -- 4 \AA\ wavelength bins
(1/10 of the resolution in the figure above), thus decreasing the
fluctuation in the effective optical depth. For reference, the
redshifted Ly$\alpha$ wavelength is at 9020 \AA, off the plot.
\label{fig:tau_sample}}
\vspace{-1\baselineskip}
\end{figure}

Since high-redshift quasars are surrounded by their own highly ionized
HII regions, it is useful to express $\tau$ as a sum of contributions
from inside ($\taures$) and outside ($\taudamp$) the \strom, $\tau =
\taures + \taudamp$.  Defined as such, the resonant optical depth,
$\taures$, would be given by Eq. (\ref{eq:total_tau}), with the {\it
lower} limit of integration changed to the redshift corresponding to
the edge of the \strom, $z(R_S)$.  The damping wing optical depth
$\taudamp$, would be given by Eq. (\ref{eq:total_tau}), with the {\it
upper} limit of integration changed to $z(R_S)$
 \footnote{Note that the terms ``resonant'' and ``damping wing'' are
only appropriate in describing the contributions of $\taures$ and
$\taudamp$ inside the \strom\ (i.e. at $\lobs \gsim \lambda_\alpha
[1+z(R_S)]$).  Outside of the \strom\ ($\lobs \lsim \lambda_\alpha
[1+z(R_S)]$), $\taudamp$ actually integrates over the resonant part of
the absorption cross section and $\taures$ integrates over the damping
wing (see Fig. \ref{fig:tau_sample}).}.

Since cosmic hydrogen is highly ionized at low redshifts, one can
replace the lower limit of integration in Eq. (\ref{eq:total_tau})
with $z_{\rm end}$, denoting the redshift below which HI absorption
becomes insignificant along the LOS to the source.  In our analysis,
we chose the value of $z_{\rm end}$ separately for each quasar to
correspond approximately to the blue edge of its GP trough.  This
roughly translates to $z_{\rm end} \sim$ 0.2--0.3 for our three
quasars.
We find that the exact choice of $z_{\rm end}$ is not important,
because most of the contribution to $\taudamp$ comes from neutral
hydrogen close to $z(R_S)$. Inside the \strom, we assume an IGM temperature of $T=2\times10^4$ K. Outside the \strom, for the case of a
mostly ionized universe, we assume an IGM temperature of $T = 10^4$ K,
while the temperature in a neutral universe was taken to be $T = 2.73
\times 151 (\frac{1+z}{151} )^2$ K, valid for $z<150$
\citep{Peebles93}.  Our results are fairly insensitive to the exact
temperature used.  The Doppler width of the Lyman-$\alpha$ absorption cross
section scales as $\nu_D \propto T^{1/2}$, but the total integrated
area under the cross section is independent of temperature.

Assuming ionization equilibrium, we compute the neutral hydrogen
fraction at each point along the LOS with:

\begin{equation}
\label{eq:nf}
n \nf(\Gamma_{\rm Q} + \Gamma_{\rm BG}) = n^2 (1-\nf)^2 \alpha_B ~ ,
\end{equation}

\noindent where $\Gamma_{\rm Q}$ and $\Gamma_{\rm BG}$ are the number
of ionizations per second per hydrogen atom due to the quasar's and
the background's ionizing fluxes, respectively, and $\alpha_B$ is the
case-B hydrogen recombination coefficient.  The LHS of Eq. (\ref{eq:nf}) accounts
for the number of hydrogen ionizations per cm$^3$, while the RHS
accounts for the number of hydrogen recombinations per cm$^3$.
Solving Eq. (\ref{eq:nf}) for $\nf$ one obtains:

\begin{equation}
\label{eq:nfsolve}
\nf = \frac{- b - \sqrt{b^2 - 4}}{2} ~ ,
\end{equation}

\noindent where $b\equiv -2 - (\Gamma_{\rm Q} + \Gamma_{\rm BG})/(n \alpha_B)$.
We further parametrize the quasar's ionization rate with:

\begin{equation}
\label{eq:fgamma}
\Gamma_{\rm Q} \equiv f_\Gamma \Gamma^e_{\rm Q}
\end{equation}

\noindent where $\Gamma^e_{\rm Q} = (4 \pi r^2)^{-1}
\int_{\nu_H}^{\infty} 1.55 \times 10^{31}$ $(\nu/\nu_H)^{-1.8}$
$[(1+z)/(1+\zsource)]^{-0.8} (\sigma/h\nu) d\nu$ s$^{-1}$, with
$\nu_H=3.29\times10^{15}$ Hz being the ionization frequency of
hydrogen and $r$ being the luminosity distance between the source and
redshift $z$.  $\Gamma^e_{\rm Q}$ results from redshifting a
power--law spectrum with a slope of $\nu^{-1.8}$, normalized such that
the emission rate of ionizing photons per second is $1.3 \times
10^{57}$, matching the emissivity one obtains \citep{HC02} for SDSS
J1030+0524 using a standard template spectrum \citep{Elvis94}. Note
that the normalization inferred from the more recent template of
\citet{Telfer02} is a factor of of $\sim5$ higher.  This roughly
translates to an emission rate of ionizing photons of

\begin{equation}
\label{eq:nion}
\dot N_Q=f_\Gamma\times1.3\times 10^{57}~{\rm s^{-1}} ~ .
\end{equation}

We ignore radiative transfer along the LOS (and also the ionization of
helium).  Instead, outside of the quasar's \strom, we compute $\nf$
assuming the quasar contributes nothing to the ionizing flux,
i.e. $\Gamma_{\rm Q}(R>R_S) = 0$, while inside the \strom, we assume
the gas is optically thin in the ionizing continuum.  This assumption
of a sharp \strom\ boundary is an approximation, valid in the regime
where either the quasar's spectrum is soft, or where a strong soft
ionizing background flux dominates over that of the quasar already at
radii smaller that the mean free path of the typical ionizing quasar
photon (as is the case of the proximity effect at lower redshift;
\citealt{BDO88}) If the quasar has a hard spectrum, and the background
intensity is low, then the edge of the \strom\ will be ``blurred''.  A
similar blurring would arise if additional ionizing sources inside the
\strom, with an extended spatial distribution, would contribute
significant flux \citep{WL06}, a possibility we do not address here.
However, already strong upper limits can be placed on the extent of
such blurring from the offset between the wavelengths of the
Ly$\alpha$ and Ly$\beta$ GP troughs.  Since Ly$\beta$ absorption is
less sensitive than Ly$\alpha$, the amount of offset between the GP
troughs gives a measure of the spatial (i.e. wavelength)
 evolution of $\tau$ close to the edge of the \strom, with a
large offset indicating slow $\tau$ evolution, and a small offset
indication rapid $\tau$ evolution. \citet{MH04} studied this for the
case of \qnametwoeight, but even stronger constraints could be placed
on the spectra of \qnamefourtwo\ and \qnametwotwo, since they do not
show any offsets between the GP troughs (Kramer et al., in
preparation).  Furhtermore, neglecting possible obscuration effects arising from optically thick gas and dust close to the quasar can bias our determination of the quasar's luminosity to lower values.  This means that our determination of the quasar's luminosity should be viewed as the observed luminosity along the LOS, and equal to the intrinsic luminosity in the low opacity limit.

Finally, we define the neutral hydrogen fraction in the IGM, $\nfIGM$,
with Eq. (\ref{eq:nfsolve}) using the mean density of the universe at
the edge of the \strom, $n = n_0 [1+z(R_S)]^3$, and $\Gamma_{\rm
Q}=0$.  Although $\Gamma_{\rm BG}$ is the more physically relevant
quantity and we use it in our analysis, henceforth we will use
$\nfIGM$ as the proxy for $\Gamma_{\rm BG}$ in order to match
convention.  As defined above, $\nfIGM$ represents the neutral
fraction at the mean density.  The more common representation of the
neutral fraction in the literature is in the form of the volume (or
mass) weighted mean, $\volavenf$. This definition, of course, depends
on the assumed density distribution.  Hence, for comparison purposes,
we express our results both in terms of $\nfIGM$ and $\volavenf$, with
the latter obtained by averaging over our simulation box densities and
assuming ionization equilibrium.

To summarize, our analysis has three free parameters:
\begin{enumerate}
\item $R_S$, the radius of the quasar's \strom,
\item $\nfIGM$, the neutral hydrogen fraction in the IGM at mean density,
\item $f_\Gamma$, the quasar's ionizing luminosity, in units of $1.3\times 10^{57}~{\rm s^{-1}}$.
\end{enumerate}
To obtain an intuition for the effect of varying these free
parameters, in Figure~\ref{fig:tau_sample} we plot $\taures$ ({\it
solid curve}) and $\taudamp$ ({\it dashed curve}) for a typical
simulated LOS, for parameter choices ($R_S$, $\nfIGM$, $\fgamma$) =
(40 Mpc, 1, 1).  As stated above, the total \lya\ optical depth is the
sum of these two contributions, $\tau$ = $\taures$ + $\taudamp$.
Roughly speaking, changing $R_S$ moves the dashed ($\taudamp$) curve
left and right, while changing $\nfIGM$ moves this curve up and down
in Figure~\ref{fig:tau_sample}. Changing $f_\Gamma$ approximately
corresponds to moving the solid ($\taures$) curve up and down.  The
dotted line demarcates roughly the maximum total optical depth, $\tau
\sim 5.5$, at which a signal can be detected with a confidence greater
than 1-$\sigma$, in the spectral region just blueward of the
Ly$\alpha$ emission line (see discussion in \S~\ref{sec:comp}).
Finally, after the optical depth and the associated transmission
($e^{-\tau}$) is calculated according to the procedure outlined above,
the transmission is averaged over $\approx$ 3.3 \AA\ wavelength bins
to match the Keck ESI spectra with which the simulated spectra are
compared (note that the actual smoothing length $\Delta\lambda_{\rm
obs}$ differs by a few percent for each source, since they are at
different redshifts).

The free parameter $R_S$ was varied in linear increments of $\Delta
R_S = 2$ Mpc, with the lower limit set by the red edge of the
Ly$\alpha$ GP through in the observed spectra.  The other parameters
were varied in the ranges: $3.2\times10^{-4} \leq \nfIGM \leq 1$, in
multiples of 5, and $0.1 \leq f_\Gamma \leq 10$, in linear increments
of $\Delta f_\Gamma = 0.3$ between $0.1 \leq f_\Gamma \leq 2$ and
$\Delta f_\Gamma = 2$ between $2 < f_\Gamma \leq 10$.

\subsection{Modeling the Intrinsic Emission Spectrum}
\label{sec:emission}

As observational input in our analysis, we make use of Keck ESI
spectra of the three highest redshift quasars discovered to date:
\qnamefourtwo\ ($\zsource=6.42$), \qnametwoeight\ ($\zsource=6.28$),
and \qnametwotwo\ ($\zsource=6.22$).  These quasars are the only
quasars discovered to date with $\zsource > 6.2$, as well as the only
quasars (aside from the BAL quasar \qnametwo) exhibiting complete GP
troughs, with no detectable flux over a wide wavelength range \citep{Fan06_disc}.  Readers interested in the details of the
corresponding observations are encouraged to consult
\citet{Fan06_disc} and \citet{White03}.

In order to compare the quasars' observed spectra with our simulated
spectra, we must know the intrinsic emission spectrum from each of the
observed sources.  Observations of lower redshift quasars
(e.g. \citealt{Telfer02}) suggest that their intrinsic continuum
emission is well fit by a broken power--law, with a mean spectral
slope of $\alpha \sim -0.7$ ($f_\nu \propto \nu^\alpha$) redward of
the Ly$\alpha$ line, and a slope of $\alpha \sim -1.8$ blueward of the
Ly$\alpha$ line.  The high-redshift SDSS quasars do not show any
evidence for deviating from this trend (e.g. \citealt{White03}).  The
precise location of the wavelength of the break in the power--law is
difficult to determine for any given source, due to a strong broad
Ly$\alpha$ emission line superimposed on the continuum spectrum.
However, for our purposes, we are interested in the optical depth at
wavelengths within just a few tens of rest-frame Angstroms blueward of
the Ly$\alpha$ line center, where the Ly$\alpha$ emission line is
still dominant over the continuum.  Therefore, we do not model the
break, and use a single power--law continuum component instead in our
emission template.  Throughout our analysis, we keep the logarithmic
slope $\alpha=-0.7$ fixed (as we will demonstrate below, our results
are insensitive to this choice).

\begin{figure*}
\begin{center}
\includegraphics[width=0.3\textwidth]{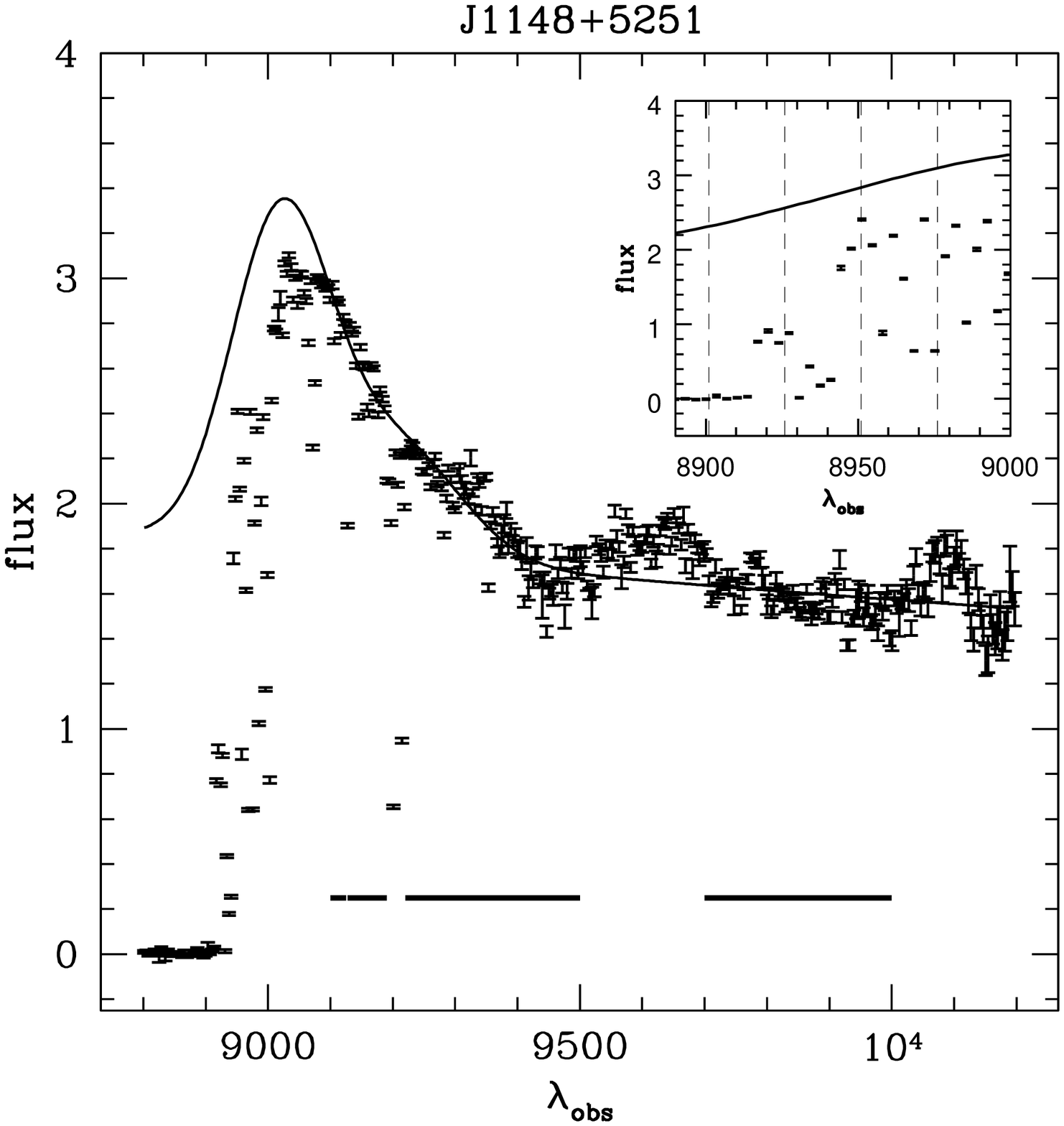}
\includegraphics[width=0.3\textwidth]{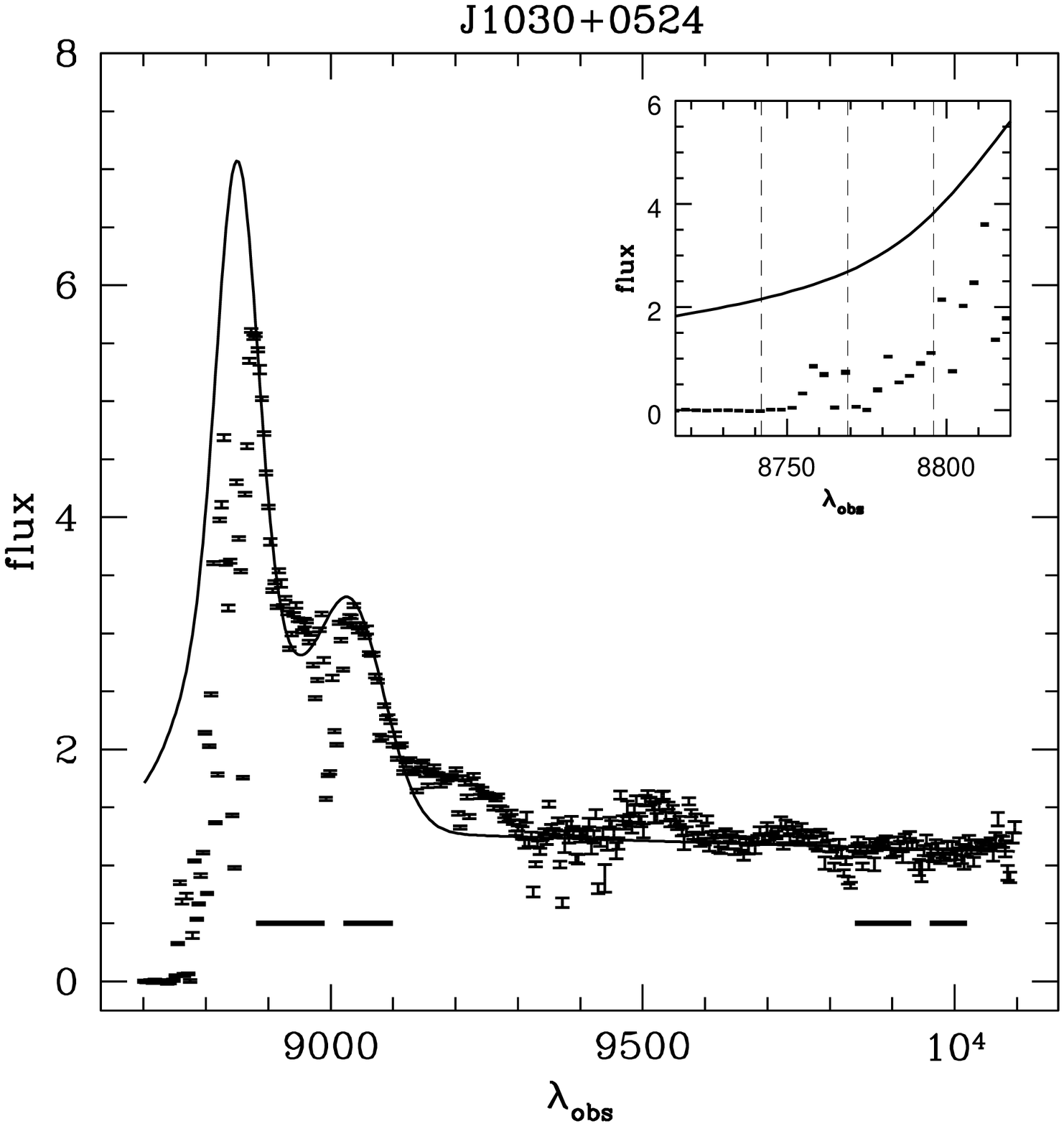}
\includegraphics[width=0.3\textwidth]{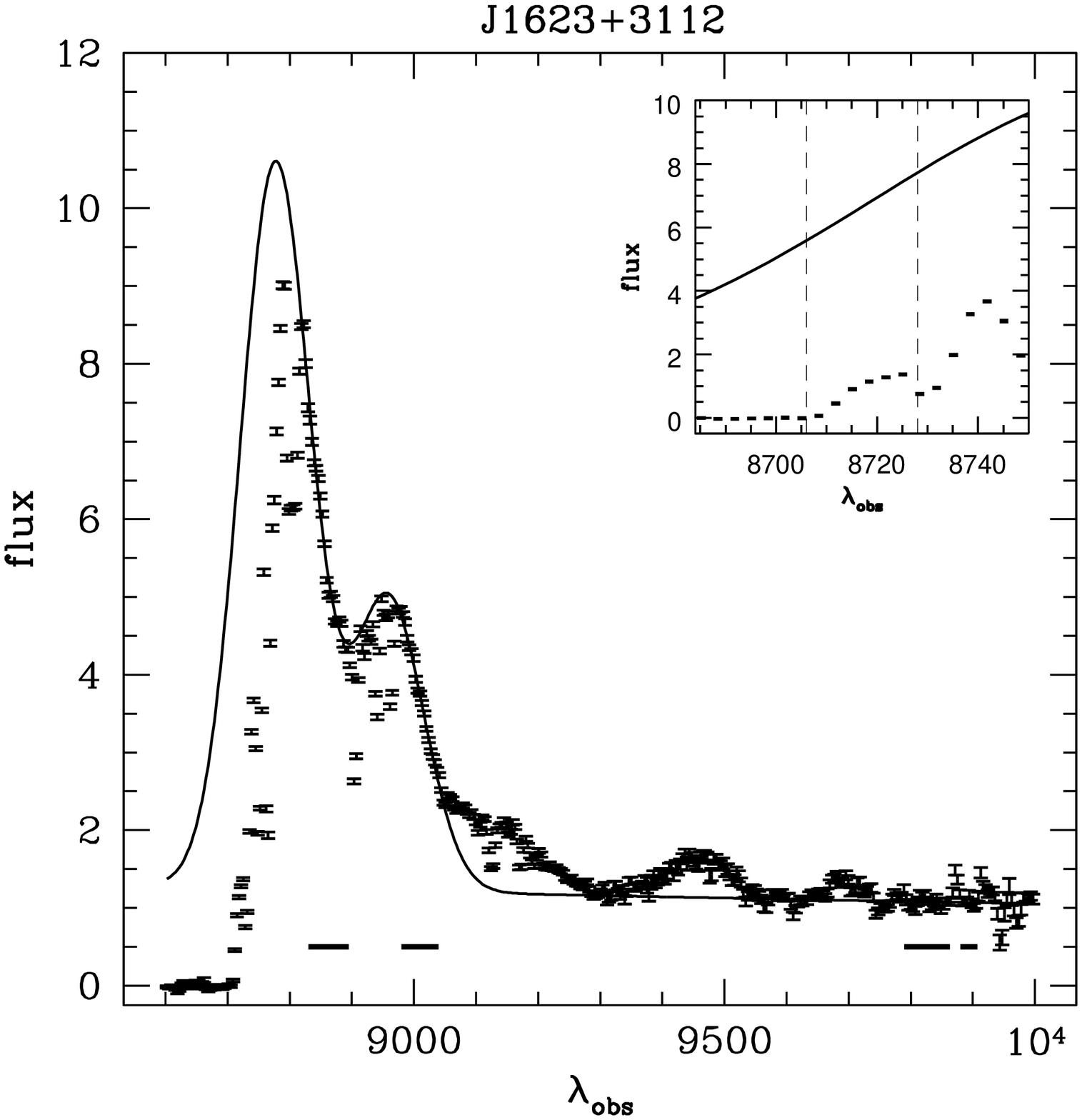}
\end{center}
\figcaption{Observed spectra (shown as points
with 1-$\sigma$ error bars) and the continuum + Ly$\alpha$ + N V fit
of the {\it intrinsic} emission spectrum generated by our procedure
({\it solid curve}).  The spectral regions (on the red side of the
Ly$\alpha$ line center) that we used in the fit are marked with
horizontal lines.  The inlaid panels zoom in on the spectral region
(on the blue side of the Ly$\alpha$ line center) that were then used
to compare the absorption optical depth $\tau$ with hydrodynamical
simulations. The vertical dashed lines bracket the wavelength ranges
used to construct $\tau$ histograms (see \S~\ref{sec:comp}).  The
spectra shown are those of \qnamefourtwo, \qnametwoeight,
\qnametwotwo\ ({\it left to right}).
\label{fig:fits}
}
\end{figure*}

We derive our constraints below entirely from the blue side of the
Ly$\alpha$ line in our analysis, where the effects of the IGM can be
felt most strongly \citep{MHC04}.  To this end, we make use of
``clean'' spectral regions on the red side of the Ly$\alpha$ line,
where the resonant attenuation is negligible ($\taures \sim 0$), and
the damping wing attenuation is less than 10\% ($e^{-\taudamp} > 0.9$)
even in the case of a fully neutral IGM and a small $R_S$.  We fit the
NV emission line at $(1+\zsource)1240.81$ \AA\ with a single gaussian,
and the continuum with a single power--law.  Additionally, we fit the
Ly$\alpha$ line with a single gaussian for \qnamefourtwo\ (with a
fitted width of $\Delta \lobs = 113$\AA) and \qnametwotwo\ (with a
fitted width of $\Delta \lobs = 81$\AA) and with a sum of two
independent gaussians for \qnametwoeight\ (with fitted widths of
$\Delta \lobs = 45$\AA\ and $\Delta \lobs = 113$\AA).  We have found
that adding a second gaussian does not improve the fits to the
Ly$\alpha$ line of \qnamefourtwo\ and \qnametwotwo. However, the
Ly$\alpha$ line of \qnametwoeight\ requires two gaussians for an
accurate fit, as it contains a broad and narrow component, similar to
the shape of a Voigt profile \citep{Ho97, GH04}.  The redshifted line
centers of the gaussians were held fixed in the fitting procedure, set
to $(1+\zsource)\lambda_0$, where $\lambda_0 = 1215.67$\AA\ for
Ly$\alpha$ and $\lambda_0 = 1240.81$\AA\ for N V.  The height and
width of the gaussians were then found by fitting the flux in pixels
on the red side of the line center.  Care was taken, in addition, to
excise the spectral region immediately redward of the Ly$\alpha$ line
center, as this region could be susceptible to non-negligible
absorption by dense infalling gas in the foreground, close to the
quasar (e.g. \citealt{MHC04}; \citealt{BL04_grbvsqso}).  Indeed, the
quasar spectra do exhibit sharp drops in the observed flux as one
approaches the line center from the red side (see Fig. \ref{fig:fits};
see also \citealt{BL03}).  This region was excised by eye from the
fitting procedure.  The amplitude of the continuum was found
separately, by fitting the flux in regions of the spectrum further to
the red, that are a-priori known not to suffer metal line absorption,
corresponding to roughly 1300 -- 1350 \AA\ in the rest frame.  We
explicitly avoid spectral sub--regions with obvious emission or
absorption lines.  More specifically, the spectral regions in the
observed frame used in the fit for the continuum were (demarcated with
horizontal lines in Figure~\ref{fig:fits}):
\begin{itemize}
\item {\qnamefourtwo:} 9100--9125\AA,  9128--9190\AA, 9220--9500\AA, 9700--10000\AA;
\item {\qnametwoeight:} 8880--8990\AA, 9020--9100\AA, 9840--9930\AA, 9960--10020\AA;
\item {\qnametwotwo:} 8830--8895\AA, 8980--9040\AA, 9790--9863\AA, 9880--9907\AA;
\end{itemize}

In Figure \ref{fig:fits}, we show the data for the observed spectra,
$F(\lobs)$, with 1-$\sigma$ error bars as well as the continuum +
Ly$\alpha$ + N V fit to the intrinsic emission, $F_0(\lobs)$,
generated by our procedure ({\it solid curve}).  The Ly$\alpha$
optical depth can then be inferred in each pixel separately as:
$\tauobs(\lobs) = - \ln[F(\lobs)/F_0(\lobs)]$.

\subsection{Comparing Simulated and Observed Spectra}
\label{sec:comp}

After the observed optical depths, $\tauobs(\lobs)$, are obtained as
discussed in \S~\ref{sec:emission}, and the simulated optical depths,
$\tausim(\lobs)$, are obtained for every LOS and point in our
parameter space as discussed in \S~\ref{sec:sim}, we wish to
statistically compare the simulated and observed optical depths for
each quasar.  More precisely, for each quasar and every point in our
three--dimensional parameter space, ($R_S$, $\nfIGM$, $f_\Gamma$), we
want to answer the question: {\it what is the likelihood that the list
of $\tauobs(\lobs)$ values and the list of $\tausim(\lobs)$ values
were drawn from the same underlying probability distribution?}  Note
that we have a fixed list of about 30 $\tauobs(\lobs)$ values for each
quasar, but we generate a different list of 3000 $\tausim(\lobs)$
values, combining 100 LOSs, for each choice of ($R_S$, $\nfIGM$,
$f_\Gamma$).

We use the Kolmogorov-Smirnov (K-S) test (e.g. \citealt{NR}) to
compare the cumulative probability distributions (CPDFs) -- i.e., histograms -- of
$\tauobs(\lobs)$ and $\tausim(\lobs)$.  The usefulness of the K-S test
is that it provides a figure of merit which can be directly
interpreted as the likelihood that two distributions were drawn from
the same underlying distribution, without the need for the ad-hoc
binning of data, and the resulting loss of information, required by
such methods as the $\chi^2$ test.  The drawback of using the K-S test
is that computing absolute confidence levels around our best fit
values would require prohibitively expensive Monte-Carlo simulations.
We caution the reader that, as a result, the relative likelihoods we
quote below do {\it not} represent traditional ``error--bars''.

Our statistical analysis is quite similar to that employed by \citet{Rollinde05} to model density structure surrounding moderate redshift ($z\sim2$) quasars. The main difference in technique is that while our present approach uses a parameterized model, \citet{Rollinde05} fit for the intrinsic spectrum directly from the low--$\tau$ regions of the spectrum.  Using simulated spectra, they show that by applying the K-S test on optical depth CPDFs, one can statistically discriminate among different models of the distance--evolution of the optical depth.  In \citet{MHC04}, we use a similar statistical test to show that the moderate degeneracy between $R_S$ and $\nfIGM$ can be broken with a single spectrum, if $R_S$ is moderately constrained.  This prior analysis did not take into account the additional constraint arising from the evolution of the CPDF with distance, as our current analysis does. 
  Nevertheless, we caution the reader that these tests only confirm the ability to statistically recover input parameters using our statistical analysis.  They do not test the validity of our model.

The optical depth CPDF should have a strong dependence on wavelength
(since the gas is more highly ionized near the quasar). In principle,
the information contained in this dependence could be extracted and
used to further constrain model parameters. For example, one could
compare the CPDFs of $\tauobs(\lobs)$ and $\tausim(\lobs)$ in several
wavelength bins for each quasar.  Unfortunately, in practice, one also
needs to have many points in each individual bin, to be able to sample
the underlying distribution in that bin.  In one extreme, as the
wavelength bin size approaches the pixel width, the information from
the wavelength dependence is fully preserved; however, only a single
observed pixel is available to compare with a corresponding pixel from
each LOS, rendering the use of the K-S test unreliable.
The converse is true in the other extreme, where the wavelength bin
size approaches the entire spectral region we analyze: the shape of the
CPDF is accurately captured, but information from the wavelength--dependence
is destroyed.  Here we
(somewhat arbitrarily) divide the spectral region of analysis into
wavelength bins of width $\sim$ 25--30 \AA, so that each bin contains
between 6--8 pixels.  The two blue--most bins were chosen bracket the
edge of the GP trough in the observed spectrum.  Furthermore, we remove the region 20--30 \AA\
blueward of the line center from our analysis, corresponding to the
expected mean radius of the large--scale overdense region surrounding such
quasars \citep{BL04_grbvsqso}.  This step was necessary, as our
simulation box size is too small to contain a statistical sample (or
even a single occurrence) of such large overdensities, required for
their accurate modeling.  The wavelength bins we use in our analysis
are shown by the vertical dashed lines in the inlaid panels of
Figure~\ref{fig:fits}.  We verify that our results are fairly
insensitive to our choice of bin size, i.e. the relative K-S likelihoods
between models change little with decreasing bin size.

In order to incorporate an uncertainty into our emission template, we
add Gaussian-distributed, uncorrelated flux variations to each pixel
in the template.  We do this by regarding our fitted value of the
template, $F_0(\lobs)$, as the mean value of the true flux
distribution, and the ``true'' emission at $\lobs$ is drawn from a
Gaussian distribution around that mean, with a standard deviation
$\sigma_{flux}(\lobs)$.  Hence, we replace every $\tausim(\lobs)$ value
by a set of 1000 values, defined by

\begin{equation}
\label{eq:sample}
\tausim^{\rm sampled}(\lobs) \equiv -\ln[e^{-\tausim(\lobs)} \times A] ~ ,
\end{equation}

\noindent where $A$ represents our emission uncertainty, and is drawn
from a Gaussian distribution with a mean of 1, and a standard
deviation of $\sigma_{flux}(\lobs)/F_0(\lobs)$.  We set
$\sigma_{flux}(\lobs)/F_0(\lobs) = 0.3$, which corresponds
approximately to the typical quasar-to-quasar scatter in the continuum
level immediately redward of \lya\ \citep{Vanden-Berk01}. However, we
have verified that our results are not sensitive to this choice.  For
example, in the case of \qnametwotwo, we find that in the range $0.1
\lsim \sigma_{flux}(\lobs)/F_0(\lobs) \lsim 0.5$, the peak K-S
probability remains unchanged, and the confidence limits quoted below
change at most by a single pixel in our parameter space.

Finally, in comparing spectra, one needs to define what one considers
to be ``zero flux'', as simulations have near-infinite precision and
observations do not.  To this end, we define a maximum allowed value
of the optical depth, ${\rm MAX}[\tau] \equiv 5.5$.  This
approximately corresponds to a 1-$\sigma$ detection in our observed
spectra at wavelengths around the edge of the GP trough, where the
majority of pixels with fluxes less than this value are located (see
Fig. \ref{fig:fits}).  In our comparison, all values of $\tau > 5.5$
are treated the same, regardless of their actual value.

To summarize, for each quasar we divide the spectral region roughly
blueward of $\sim$ $(1+\zsource)\lambda_\alpha$--25\AA\ and redward of
$\sim$ $[1+z(R_S)]\lambda_\alpha$--25\AA\ into 3--5 bins of approximately equal
wavelength width, containing 6--8 pixels each.  In each bin we compare the CPDFs of
$\tauobs(\lobs)$ and $\tausim^{\rm sample}(\lobs)$ (the latter
concatenated over 100 simulated LOSs).  For every point in our 3D
parameter space, these two distributions are compared via the K-S
test, which produces a probability to be interpreted as the likelihood
that the two distributions were drawn from the same underlying
distribution.  Finally, the K-S probabilities for each wavelength bin
are multiplied to give a likelihood estimate for that point in the 3D
parameter space.

\section{Results}
\label{sec:results}

\begin{figure*}
\begin{center}
\includegraphics[width=0.3\textwidth]{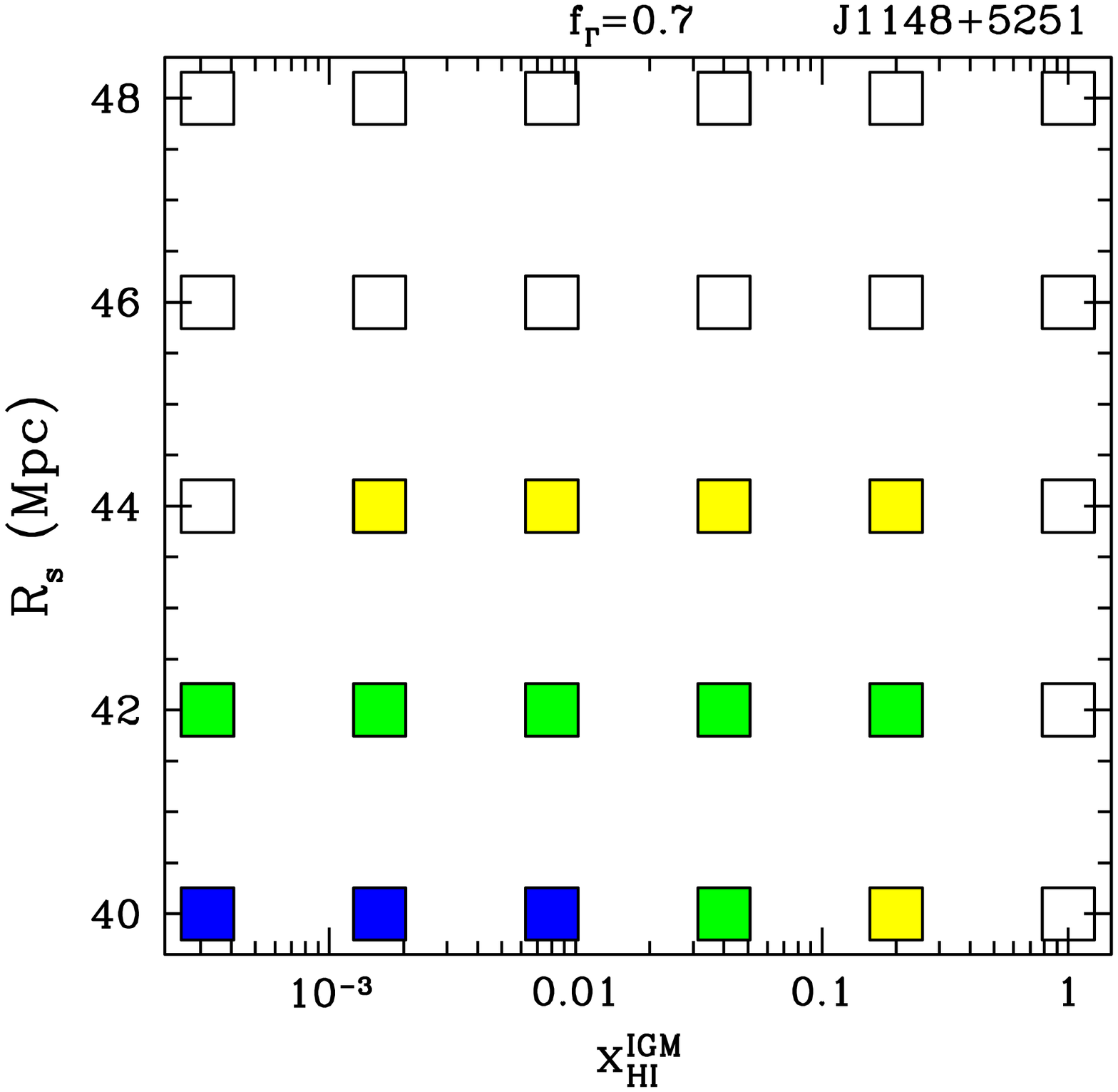}
\includegraphics[width=0.3\textwidth]{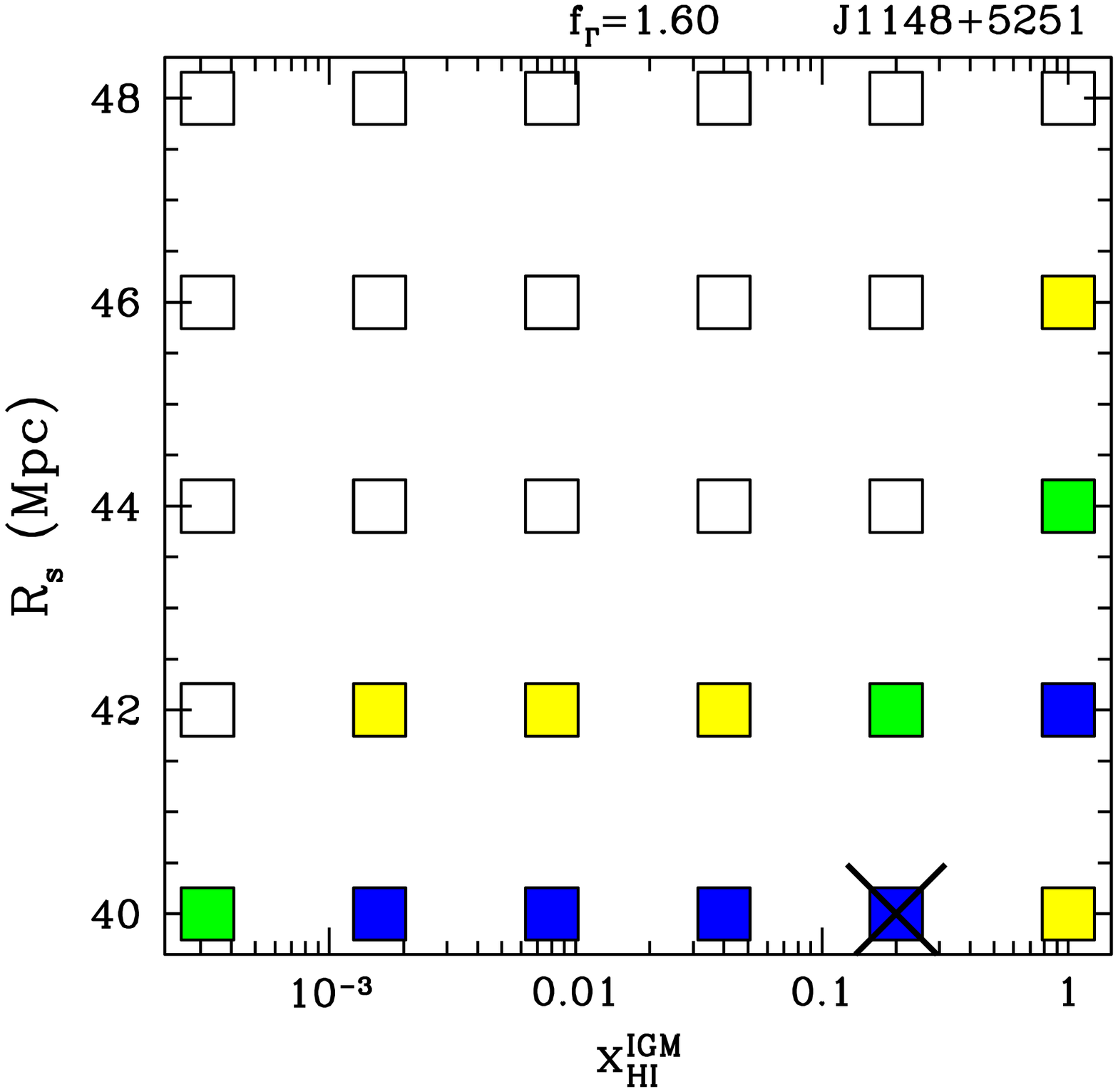}
\includegraphics[width=0.3\textwidth]{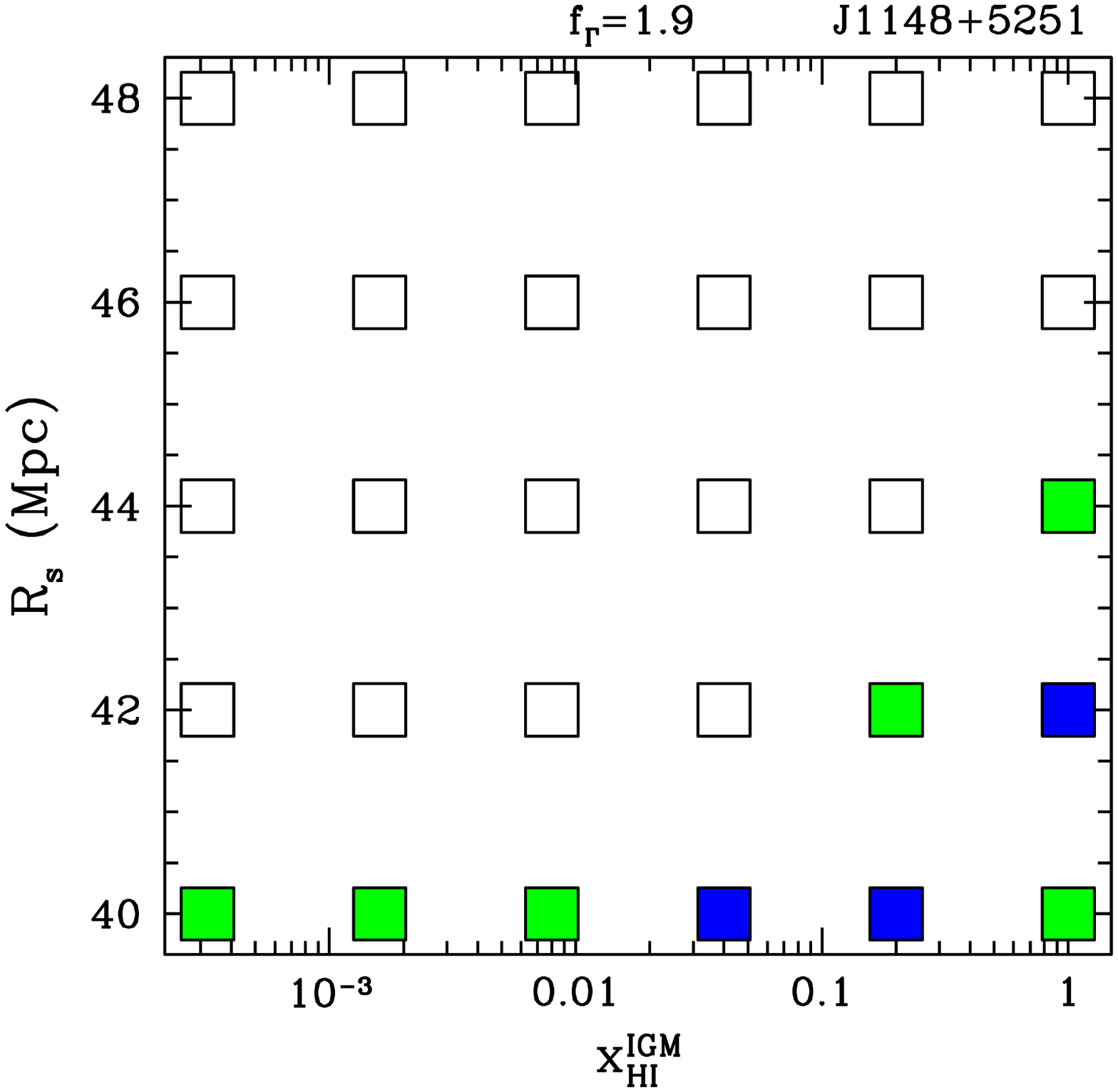}

\includegraphics[width=0.3\textwidth]{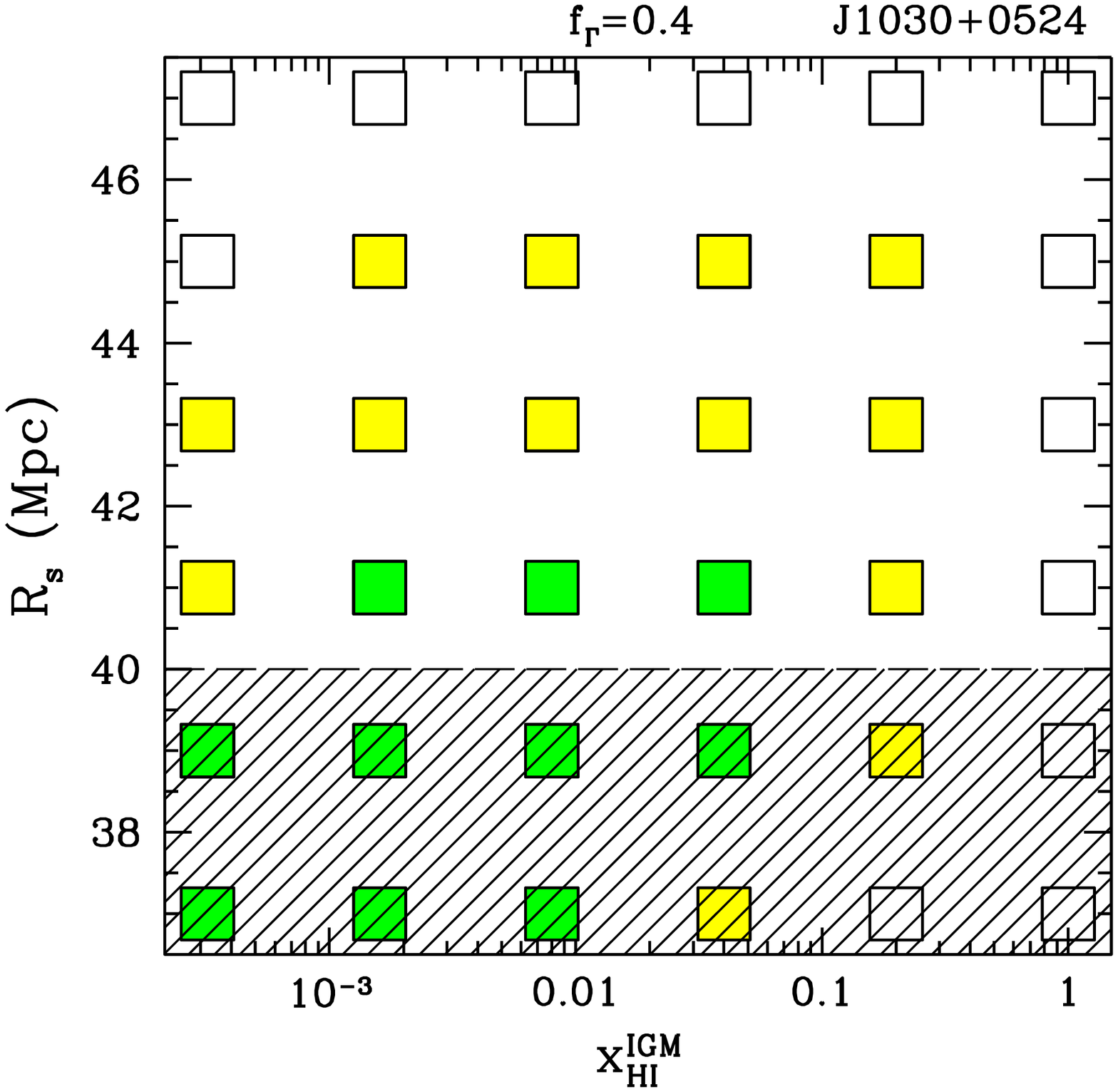}
\includegraphics[width=0.3\textwidth]{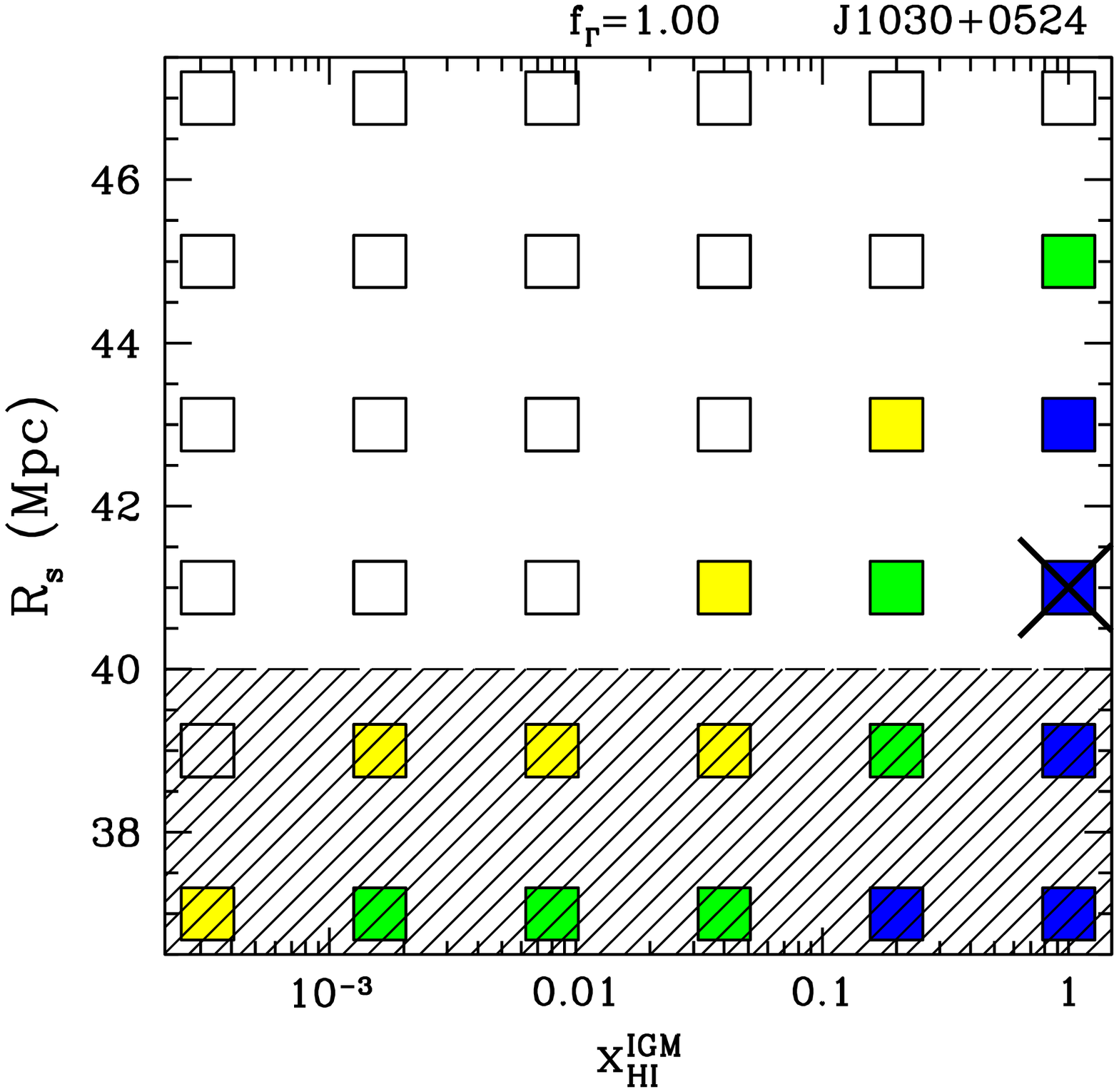}
\includegraphics[width=0.3\textwidth]{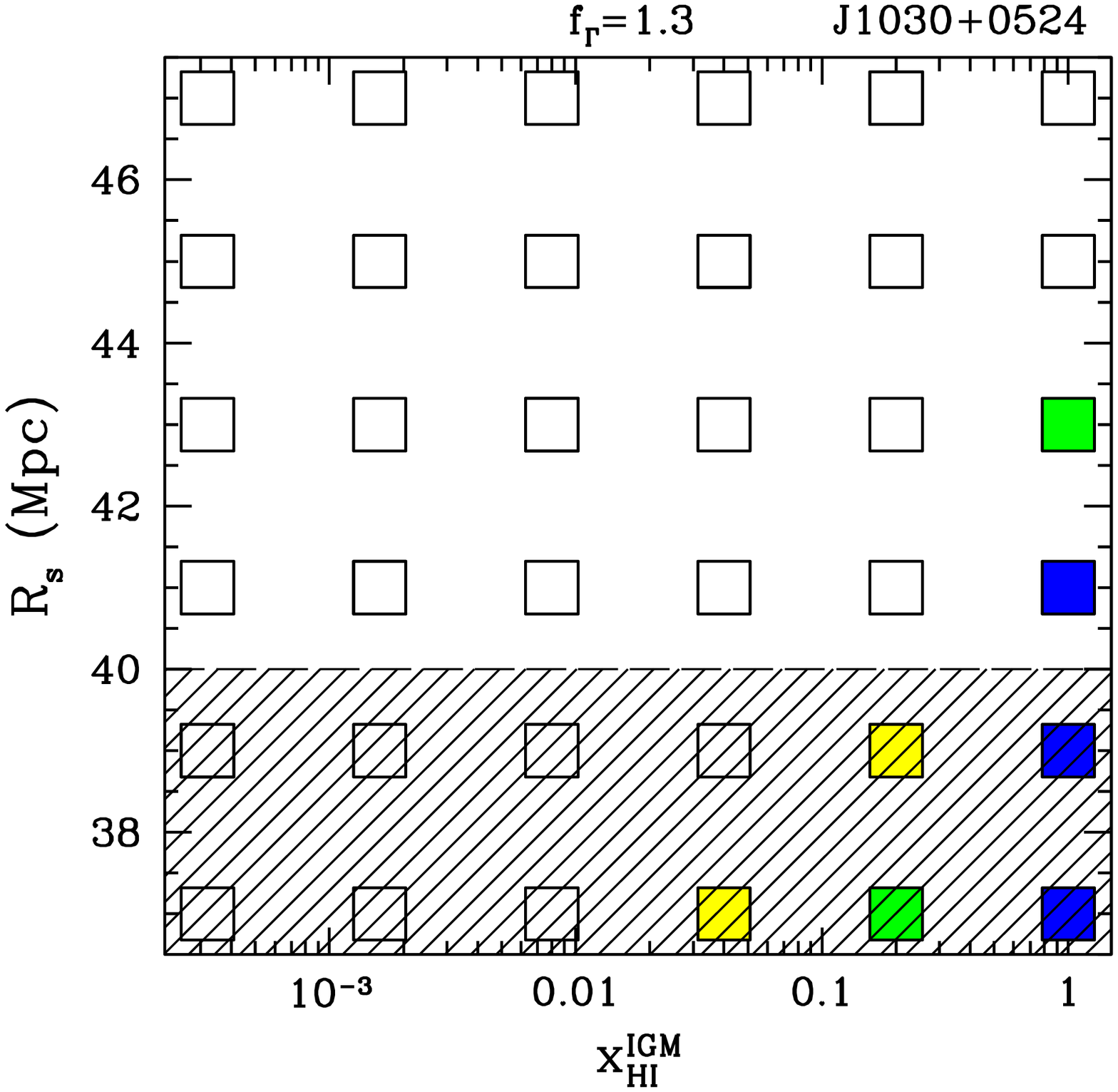}

\includegraphics[width=0.3\textwidth]{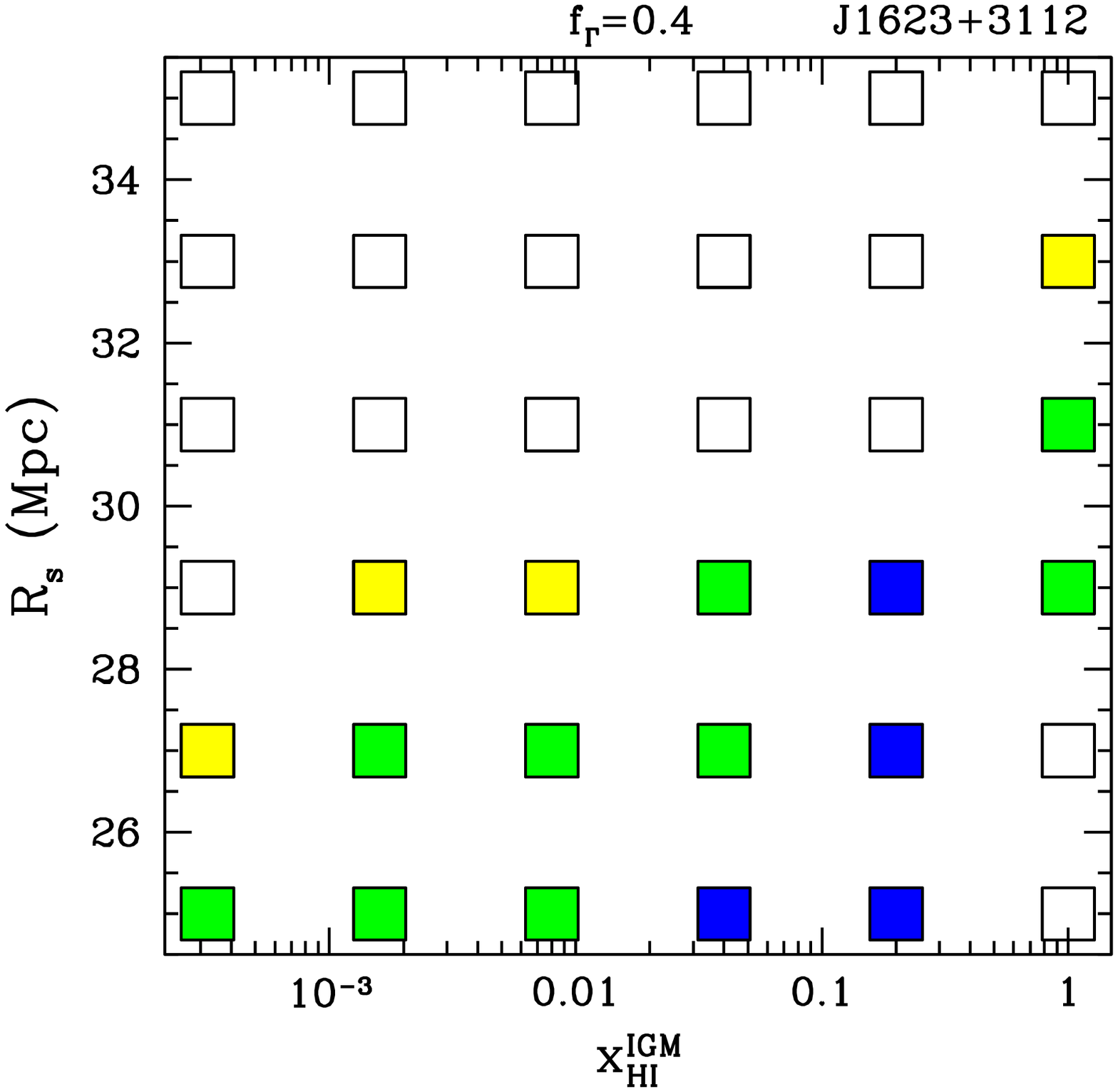}
\includegraphics[width=0.3\textwidth]{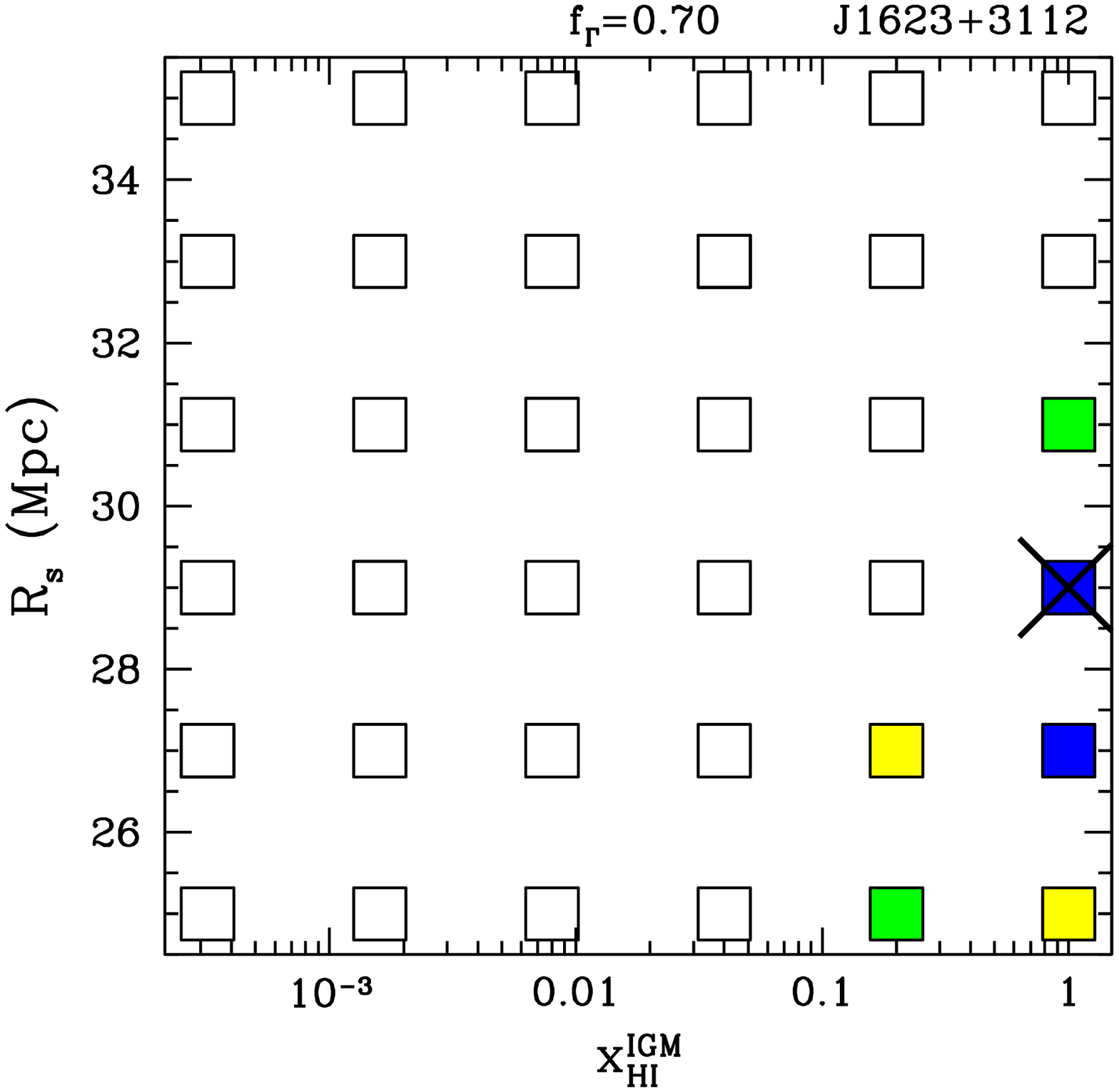}
\includegraphics[width=0.3\textwidth]{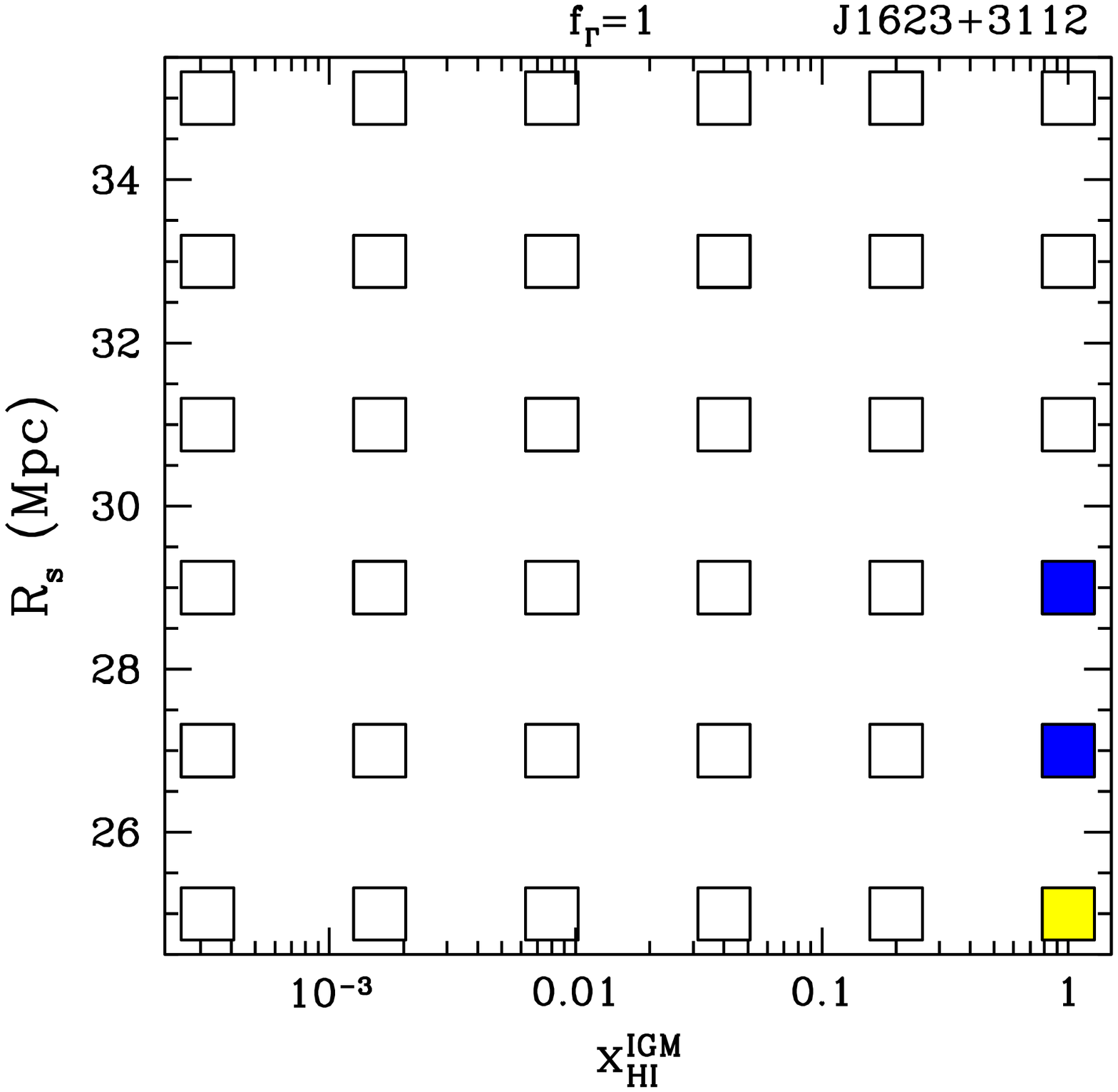}
\end{center}
  \figcaption{Likelihoods for \qnamefourtwo\
  ({\it top}), \qnametwoeight\ ({\it middle}), and \qnametwotwo\ ({\it
    bottom}) in the 3--dimensional parameter space of ($R_S$,
  $\nfIGM$, $\fgamma$).  Each panel shows contours in a 2D slice of
  the 3D likelihood surface. The cross marks the parameter combination
  with the peak likelihood value (see \S~\ref{sec:comp} for how the
  likelihoods were derived). In order from lighter to darker, the
  yellow, green, and blue squares correspond to points in our
  parameter space with likelihoods that are a factor of 27, 9, and 3
  below the peak value, respectively.  Parameter combinations with
  likelihoods below 1/27th of the peak are shown as empty squares.
  {\emph \qnamefourtwo:} The peak likelihood of 3.0\% occurs at
  ($R_S$, $\nfIGM$, $\fgamma$) = (40 Mpc, 0.2, 1.6).  The left,
  center, and right panels correspond to values of $\fgamma$= 0.7,
  1.6, 1.9, respectively.  {\emph \qnametwoeight:} The peak likelihood
  value of 34\% occurs at ($R_S$, $\nfIGM$, $\fgamma$) = (41 Mpc, 1,
  1.0).  The left, center, and right panels correspond to values of
  $\fgamma$= 0.4, 1.0, 1.3, respectively.  The shaded region shows
  values ruled out by the presence of flux in the Lyman $\beta$
  region, which sets the additional constraint $R_S$ $\gsim$ 40 Mpc.
  {\emph \qnametwotwo:} The peak likelihood value of 39\% is at
  ($R_S$, $\nfIGM$, $\fgamma$) = (29 Mpc, 1, 0.7).  The left, center,
  and right panels correspond to values of $\fgamma$= 0.4, 0.7, 1.0,
  respectively.
\label{fig:qsoparams}}
\vspace{-1\baselineskip}
\end{figure*}

Our results for the likelihoods of various parameter--combinations are
shown in Figure~\ref{fig:qsoparams} for \qnamefourtwo\ ({\it top}),
\qnametwoeight\ ({\it middle}), and \qnametwotwo\ ({\it bottom}).  The
cross in each case marks the point in the 3D parameter space where the
likelihood peaks.  In order from lighter to darker, the yellow, green,
and blue squares correspond to points in our parameter space with
likelihoods that are a factor of 27, 9, and 3 below the peak
likelihood, respectively.\footnote{In order to obtain absolute
  confidence limits, we would need to perform a set of 
  computationally expensive Monte-Carlo simulations.
  For  simplicity, we  avoid  such computations,  and instead  contend
  ourselves by  quoting results in  terms of the relative  offset from
  the   peak  likelihood.}  
Parameter combinations with likelihoods smaller than 1/27th of the peak
are shown as empty squares.

Below, we will discuss each quasar in turn.  However, it is
encouraging to note first that the isocontours in Figure
\ref{fig:qsoparams} behave as expected, following the moderate
degeneracies in the parameters.  Namely, since a stronger damping wing
($\taudamp$) contribution can be obtained either by increasing the
neutral fraction or by decreasing the size of the HII region, the
isocontours should be oriented from bottom-left to upper-right in each
panel.  Similarly, since a smaller ionizing flux (i.e. greater
$\taures$) can roughly be compensated for in the total optical depth,
$\taures+\taudamp$, by increasing the size of the HII region or
decreasing the neutral fraction (both decrease $\taudamp$), the
isocontours should shift monotonically from the upper--left to the
bottom--right with increasing $\fgamma$.  Both of these trends are
evident in Figure \ref{fig:qsoparams}.  For example, the colored squares (corresponding to K-S likelihoods that are within a factor of 27 of the peak likelihood) in the middle row cover most of the displayed parameter space in the left--most panel, though there are no pixels with high likelihoods.  Increasing the quasar's ionizing flux (i.e. going through the panels from left to right), shifts the colored squares increasingly towards the bottom--right corner of the panel.

\subsection{\qnamefourtwo}
\label{sec:qso42}

Likelihood estimates for \qnamefourtwo\ ($\zsource=6.42$) can be seen
in the top row of three panels in Fig. \ref{fig:qsoparams}.  The left,
center, and right panels correspond to values of $\fgamma$= 0.7, 1.6,
and 1.9, respectively.  The peak likelihood of 3.0\% occurs at ($R_S$,
$\nfIGM$, $\fgamma$) = (40 Mpc, 0.2, 1.6).  
This is the smallest peak likelihood we obtain (c.f. peak likelihoods of $\sim$30\% for the two quasars below), and most likely means that our model isn't as accurate in describing the physical spectra of this particular quasar.
The volume--averaged
neutral fraction at the peak is $\volavenf=0.16$.  The range of
parameter values whose likelihood estimates are within a factor of 3
of the peak likelihood is encompassed by:
\begin{itemize}
\item 40 Mpc $\lsim R_S \lsim$ 42 Mpc,
\item $\nfIGM \lsim$ 1.0 or $\volavenf \lsim$ 1.0,
\item 0.7 $\lsim f_\Gamma \lsim$ 1.9.
\end{itemize}
Note that here as well as below, the end points of the quoted ranges
correspond to the minimum or maximum value of the parameter for which
there exists a combination of the other two parameters giving a K-S
probability within a factor of three from the peak probability.
Note that these ranges are similar in spirit to ``marginalized errors'', 
in that the other parameters are allowed to vary, rather than fixed (but
they are not actual marginalized errors; see caveat mentioned in 
the footnote above).

\subsection{\qnametwoeight}
\label{sec:qso28}

Likelihood estimates for \qnametwoeight\ ($\zsource=6.28$) are shown
in the middle row of panels in Figure~\ref{fig:qsoparams}.  The left,
center, and right panels correspond to values of $\fgamma$= 0.4, 1.0,
and 1.3, respectively.  The peak likelihood of 34\% occurs at ($R_S$,
$\nfIGM$, $\fgamma$) = (41 Mpc, 1, 1.0).  The range of parameter values
whose likelihood estimates are within a factor of 3 of the peak
likelihood is encompassed by:
\begin{itemize}
\item 37 Mpc $\lsim R_S \lsim$ 45 Mpc,
\item $\nfIGM \lsim$ 1.0 or $\volavenf \lsim$ 1.0,
\item 0.7 $\lsim f_\Gamma \lsim$ 1.9.
\end{itemize}

As mentioned earlier, \qnametwoeight\ has a unique (at least among our
sample of three quasars) feature.  Namely, the Ly$\beta$ GP through is
noticeably offset from the Ly$\alpha$ GP through.  The mere {\it
  presence} of flux (irrespective of its properties) in the Ly$\beta$
region of the spectrum can be used to set a minimum size of the
surrounding HII region, $R_S \gsim 40$ Mpc.  The region excluded by
this prior is shaded over in the middle row in Figure
\ref{fig:qsoparams}.  Taking this lower limit on $R_S$ into account,
the range of values whose likelihood estimates are within a
factor of 3 of the peak likelihood shrinks to:
\begin{itemize}
\item 41 Mpc $\lsim R_S \lsim$ 45 Mpc,
\item 0.04 $\lsim \nfIGM$ or  0.033 $\lsim \volavenf$,
\item 0.7 $\lsim f_\Gamma \lsim$ 1.3.
\end{itemize}

\subsection{\qnametwotwo}
\label{sec:qso22}

Likelihood estimates for \qnametwotwo\ ($\zsource=6.22$) are shown in
the bottom row of panels in Figure \ref{fig:qsoparams}.  The left,
center, and right panels correspond to values of $\fgamma$= 0.4, 0.7,
and 1.0, respectively.  The peak likelihood of 39\% occurs at ($R_S$,
$\nfIGM$, $\fgamma$) = (29 Mpc, 1, 0.7).  The range of parameter values
whose likelihood estimates are within a factor of 3 of the peak
likelihood is encompassed by:
\begin{itemize}
\item 25 Mpc $\lsim R_S \lsim$ 29 Mpc,
\item 0.04 $\lsim \nfIGM$ or  0.033 $\lsim \volavenf$,
\item 0.4 $\lsim f_\Gamma \lsim$ 1.3.
\end{itemize}

\begin{figure}
  \vspace{+0\baselineskip}
  \myputfigure{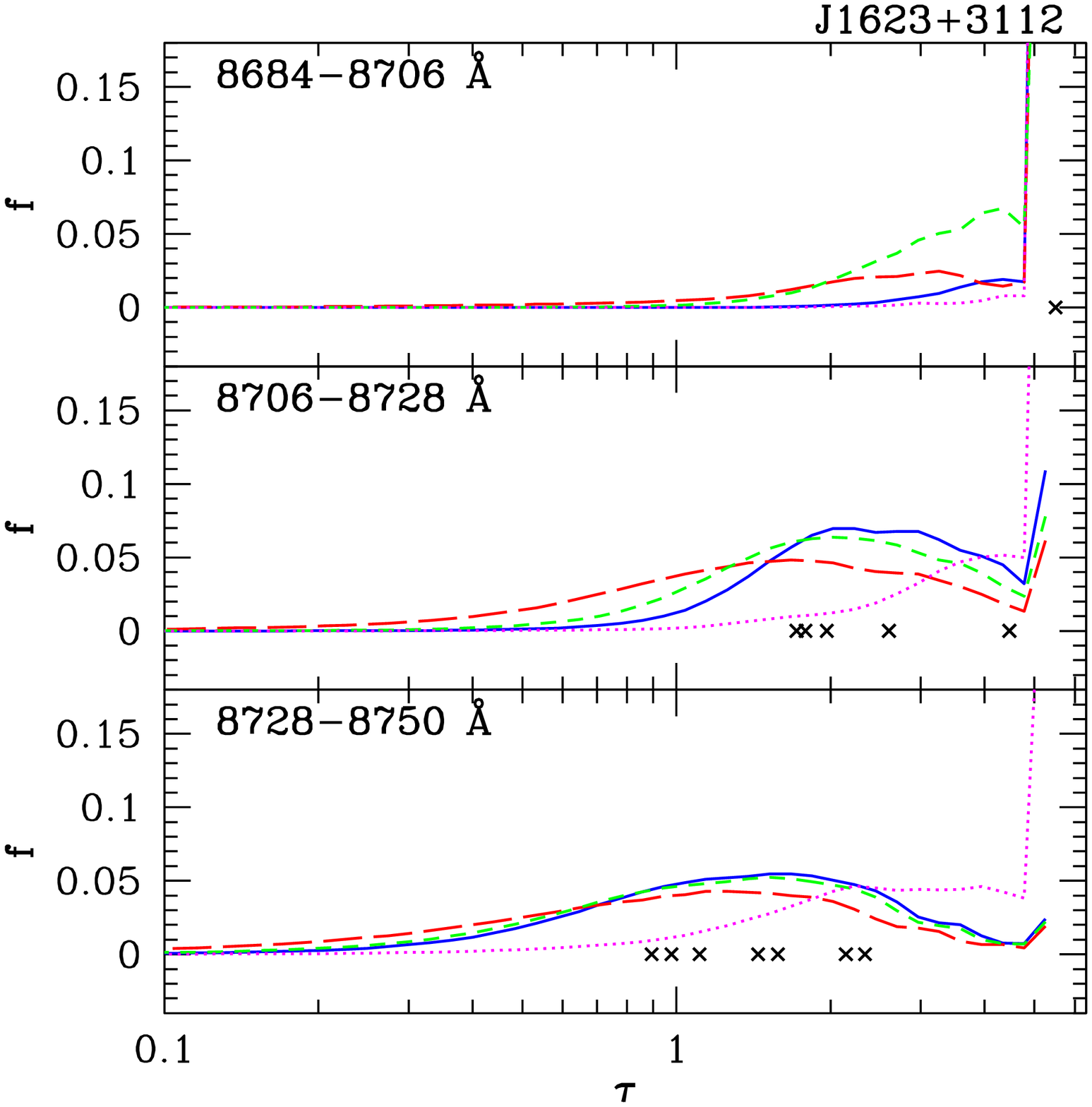}{3.3}{0.5}{.}{0.}  \figcaption{
  Distributions of simulated optical depths for the three wavelength
  bins used in our analysis of \qnametwotwo.  The solid blue,
  long-dashed red, short-dashed green, and dotted purple curves
  correspond to parameter combinations of ($R_S$, $\nfIGM$, $\fgamma$)
  = (29 Mpc, 1.0, 0.7), (29 Mpc, 0.04, 0.7), (33 Mpc, 1.0, 0.7), and
  (29 Mpc, 1.0, 0.1), respectively.  The crosses are located at the
  set of observed optical depth values.  The solid blue curve
  corresponds to the parameter combination with the peak likelihood
  value, while the parameter combinations represented by the other
  curves have likelihood values below 1/27th of the peak.  The sharp
  jump at the largest $\tau$ bin and the overlaying of seven crosses
  in the top panel are due to the finite detection threshold, whereby
  all values of $\tau > 5.5$ are indistinguishable from one another,
  and hence fall in the same bin in this figure.  The figure
  demonstrates that our lower limit on $\nfIGM$ comes from the
  low--$\tau$ tail predicted in models with low neutral fractions or
  those corresponding to regions of parameter space which could mimic
  a low neutral fraction (see the discussion on isocontours in
  \S~\ref{sec:results}).  No points corresponding to these tails are
  seen in the data.  Note also that while some models might perform
  well in one wavelength bin, only the solid blue curve is consistent
  with the observed data in {\it all} wavelength bins.
\label{fig:dist_tau}}
\vspace{-1\baselineskip}
\end{figure}

\section{Discussion}
\label{sec:discussion}

In this section, we discuss a few important aspects of the above
results.  First, it is encouraging that the simple methodology we
outlined can deliver statistically interesting constraints, even from
the few dozen available spectral pixels.  Clearly, it will be
interesting to apply a similar analysis to a larger sample of sources
in the future. 

Perhaps our most interesting result is the lower limit we obtain, for two
of the quasars, on the neutral fraction.  We would like to therefore
understand where these constraints actually come from.  To this end,
in Figure~\ref{fig:dist_tau} we show the simulated optical depth PDFs
for the three wavelength bins used in our analysis of \qnametwotwo.
The solid blue, long-dashed red, short-dashed green, and dotted purple
curves correspond to parameter combinations of ($R_S$, $\nfIGM$,
$\fgamma$) = (29 Mpc, 1.0, 0.7), (29 Mpc, 0.04, 0.7), (33 Mpc, 1.0,
0.7), and (29 Mpc, 1.0, 0.1), respectively.  The crosses are located
at the set of observed optical depth values.  The solid blue curve
corresponds to the parameter combination with the peak likelihood
value, while the parameter combinations represented by the other
curves have likelihood values below 1/27th of the peak.  The sharp
jump corresponding at the largest $\tau$ bin  and the overlaying of seven crosses in the top panel are due to the finite
detection threshold, whereby all values of $\tau > 5.5$ are
indistinguishable from one another, and hence fall in the same bin in
this figure.  The figure clearly demonstrates that our lower limit on
$\nfIGM$ arises from the low--$\tau$ tail predicted in models with low
neutral fractions or those corresponding to regions of parameter space
which could mimic a low neutral fraction (see the discussion on
isocontours in \S~\ref{sec:results}).  While the solid blue curves are
consistent with the distribution of points, these other curves all
have tails that are ``too long'': no points corresponding to these
tails are seen in the data.  Note also that while some models might
perform well in one wavelength bin, only the solid blue curve is
consistent with the observed data in {\it all} wavelength bins.

\subsection{Sanity Checks}
\label{sec:sane}

As our analysis technique is as yet untested, it is useful to take a step back, and verify that our results make sense.  As already mentioned, the
likelihood iso--contours behave as expected.  Here we discuss
several other encouraging checks on the validity of our analysis.  We caution the reader, however, that these are not meant to be a proof of the accuracy of our analysis, and instead should be regarded merely as consistency checks.

First, the procedure here recovers fairly well the result of our
previous analysis in the case of \qnametwoeight\ \citep{MH04}.  This
need not be the case, since the nature of the two analysis procedures
are different; namely, our previous work focused only on gross,
approximate Ly$\alpha$ {\it and} Ly$\beta$ transmission features near
the edge of the HII region.  Nevertheless, in both cases, the peak
likelihood occurs at the same value of $\nfIGM$ and $R_S$, with only
the value of $\fgamma$ being somewhat larger here ($\fgamma$=1.0
instead of $\fgamma$=0.4 in \citealt{MH04}).  We attribute this shift
in $\fgamma$ primarily to the inclusion of wavelength bins further
inward from the HII region edge, where the effects of $R_S$ and
$\nfIGM$ are felt less strongly and $\fgamma$ wields the most
statistical weight.  In particular, the likelihoods in the two
blue-most wavelength bins are comparable for $\fgamma=1.0$ and
$\fgamma=0.4$ (given $\nfIGM=1.0$ and $R_S=41$Mpc); however, the
likelihoods in the two red-most wavelength bins are $\sim$ 7--8 times
greater for $\fgamma=1.0$ than $\fgamma=0.4$.

Second, our results on the neutral fraction agree in relative terms
with estimates obtained from transmission in GP troughs \citep{Fan06}.
\qnametwoeight\ and \qnametwotwo\ both have very dark GP troughs in
Ly$\alpha$ and Ly$\beta$, and thus are only able to provide lower
limits on the optical depth and corresponding neutral fraction;
however, \qnamefourtwo\ has transmission gaps in the GP troughs, thus
favoring a smaller optical depth and corresponding neutral fraction.
This is qualitatively similar to our results: we were unable to
constrain $\nfIGM$ from the spectra of \qnamefourtwo\, obtaining only
a shallow peak in the likelihood at $\nfIGM=0.2$; however, both
\qnametwoeight\ and \qnametwotwo\ had peak likelihoods at $\nfIGM=1$,
and strongly prefer $\nfIGM \gsim 0.04$.

Third, the maximum likelihoods of $\fgamma$ (i.e. the relative
strength of the quasars' ionizing luminosity) match the relative
strengths of the quasars' continua redward of Ly$\alpha$.  The values
of $\fgamma$ corresponding to the peak likelihoods are 1.6, 1.0, and 0.7
for \qnamefourtwo\, \qnametwoeight\, and \qnametwotwo\, respectively.
Looking at Figure \ref{fig:fits}, one can note that this is also the
order of the relative strengths of the quasars' continua redward of
Ly$\alpha$, as one would expect if the UV power-law indices and amount of obscuration didn't
vary greatly among the quasars.

Finally, the quasar with the smallest HII region, \qnametwotwo\,
with a peak likelihood at $R_S = 29$ Mpc, also has the highest favored
value of $\nfIGM=1.0$ and the lowest favored value of $\fgamma$.  This
makes sense, since a fainter quasar, embedded in a more neutral medium
should have a smaller HII region, given a similar lifetime.

\subsection{Uncertainties in the Intrinsic Emission Spectra}

Although we have added gaussian noise to our inferred template
spectrum, as described above, we wish to get a sense of how {\it
  correlated} errors, modifying the broader shape of the template
spectrum, would affect our results.  The general concern is that
mis--estimates of the flux, correlated over many pixels, could
potentially mimic the effect of shifts in our parameter space.  To get
a sense of the magnitude of this effect, we chose to vary the width of
the fitted Ly$\alpha$ gaussian for quasar \qnametwotwo\ and see how
this modifies our results.  Naively, one would expect a narrower
Ly$\alpha$ emission line and an accompanying decrease in effective
$\tauobs$ to cause a shift in the parameter space to regions favoring
smaller optical depths (larger $R_S$, smaller $\nfIGM$, larger
$\fgamma$), with the opposite trends for a wider Ly$\alpha$ emission
line.

We find that such degeneracies are very weak.  For a 10\% narrower
line, we find that the peak likelihood remains at the same
parameter--combinations. The range of parameter values whose
likelihood estimates are within a factor of 3 of the peak likelihood
changes by at most one pixel in our 3D parameter grid: 25 Mpc $\lsim
R_S \lsim$ 29 Mpc; 0.008 $\lsim \nfIGM \lsim$ 1.0; 0.4 $\lsim f_\Gamma
\lsim$ 1.6.  This is a rather small change, and we emphasize that a
10\% narrower Ly$\alpha$ line would actually be an unacceptably
poor fit to the (red side of the) observed spectrum.
A wider Ly$\alpha$ line is more physically motivated, since strong
damping wing absorption might be able to somewhat attenuate flux far
into the red side of the line.  Hence, we vary the width of the
Ly$\alpha$ line, well in excess of the absorbing potential of a strong
damping wing.  Again, we find little change in our results.  Even for
a 50\% wider line,
the parameter combination at peak likelihood remains the same, while
the range of parameter values whose likelihood estimates are within a
factor of 3 of the peak likelihood changes by at most one pixel to: 25
Mpc $\lsim R_S \lsim$ 31 Mpc; 0.2 $\lsim \nfIGM \lsim$ 1.0; 0.4 $\lsim
f_\Gamma \lsim$ 1.3.  As expected, an intrinsically wider Ly$\alpha$
line shows a slight preference for a more neutral universe.  

Overall, the above exercise suggests that our results are robust and
conservative, at least to mis--estimates of the Ly$\alpha$ line
width. This does not, of course, preclude the possibility that the
intrinsic Ly$\alpha$ lines of the $z\approx 6$ quasars possess some
deviations from a gaussian shape.  For example, if there is an
asymmetry, in the sense of a steeper drop of flux on the blue side of
the line than on the red side, our technique would have mistakenly
attributed this steeper drop to absorption. This implies that we might
have overestimated the neutral fractions. At present, there is,
however, no evidence of such an asymmetry in the emission lines of low
redshift quasars (where both sides of the lines can be seen
unabsorbed), nor is there evidence, from the red side of the lines,
that the high--$z$ quasars' lines have shapes different from those of
their lower--redshift counterparts.

Finally, we also note that our spectral fits, describing the intrinsic
emission line on the unabsorbed red side of the line, leave typical
flux residuals of only $\sim1\%$.  Similar residuals are presumably
present on the blue side of the line, as well. However, the absorption
$\exp(-\tau_D)$ caused by the IGM damping wing on the blue side of the
line (see Fig.~\ref{fig:tau_sample}), has an amplitude of $\sim
(0.1--1)\nf$, exceeding the residuals for $\nf \gsim 0.01$.
Furthermore, in order for the residuals to mimic the IGM damping wing,
they would have to be coherent (monotonically increasing with
wavelength).

\subsection{Biases in Local Density and Velocity Fields}

As mentioned above, we remove the region 20--30 \AA\
blueward of the line center from our analysis, corresponding to the
expected mean radius of the large--scale overdense region surrounding such
quasars \citep{BL04_grbvsqso}.  This is a necessary step as present-day simulations cannot statistically model such rare, large--scale overdensities.  However, if the density bias extended into our region of analysis, our inferred $\nfIGM$ lower limits would be conservative and our inferred $\fgamma$ values would most likely be underestimated.  If the source is hosted by a large--scale overdensity, then density would decrease with decreasing $\lobs$ (increasing distance from the source).  If such a decreasing density extended into our region of analysis, it would translate into a shallower rise in $\taures \propto n^2/\Gamma$.  In matching the total optical depth $\tautot = \taures + \taudamp$, our analysis, which assumes constant mean density, would translate this shallower rise in $\taures$ to a shallower rise in $\taudamp$, i.e. a lower value of $\nfIGM$ than appropriate if the density bias was taken into account\footnote{Note that a shallower rise in $\taudamp$ could also be achieved if we were overestimating $R_S$, instead of underestimating $\nfIGM$.  However, our inferred $R_S$ values are already at the lowest limit set by the onset of GP troughs for two of our quasars and thus can not be decreased further.}.
Note that our results on $\nfIGM$ are driven by the wavelength dependence of $\tau$; a constant offset in the density field can be absorbed by a shift in $\Gamma_{\rm Q}$ (i.e. $\fgamma$), which dominates over $\Gamma_{\rm BG}$ throughout most of our region of interest and is predominantly set by the amount of absorption in bins further from the HII region edge where it has the most statistical weight.
Furthermore, it is possible that an infalling IGM could introduce a similar bias, but the effect is likely to be small for a bright quasar and far away from the
line center on the blue side.  Moreover, the effect from the infall
would have the same sign as the density bias above, making the flux fall off on the blue side of the line less steeply \citep{DHS06, BL03}).  In other words, by ignoring the possible density and velocity biases, we might only be slightly {\it under}estimating the neutral fraction.

\subsection{Uncertainties in Intrinsic Redshift}

A different source of possible error, in estimates of $R_S$, can
result from an incorrect redshift determination.  The redshifts we
quote above were determined from high-ionization lines such as C IV,
Si IV and N V.  However, high-ionization emission lines can have
non-negligible offsets from low-ionization, narrow, or molecular lines
in the host galaxy.  Recently, redshift determinations based on such
lines have been made for the three quasars used in our analysis:
$z_{\rm CO}=6.42$ for \qnamefourtwo\ \citep{Walter03}; $z_{\rm Mg
  II}=6.31$ for \qnametwoeight\ \citep{Iwamuro04}; $z_{\rm Mg II}=6.22$
for \qnametwotwo\ (L. Jiang et al. 2006, in preparation).  These
redshift determinations agree with the ones based on high-ionization
lines used in this work for \qnamefourtwo\ and \qnametwotwo; however
the Mg II redshift is 0.03 larger than the high-ionization line
redshift for \qnametwoeight.  If one takes the Mg II redshift to be
the true systematic redshift, this would roughly increase all $R_S$
determinations for \qnametwoeight\ by $\sim$ 30\% The majority of the
effect would be confined to the increase in $R_S$, since the
systematic redshift doesn't impact the Ly$\alpha$ emission fitting,
and the isocontours (as seen in the middle panels of
Fig. \ref{fig:qsoparams}) are steep (i.e. a large increase in $R_S$
can be compensated for by a small increase in $\nfIGM$, especially for
the lower confidence contours).  Since increasing $R_S$ can roughly be
compensated for by increasing $\nfIGM$ or decreasing $\fgamma$, the
secondary effects of such a redshift underestimate should be a shift
in the favored parameter values towards larger $\nfIGM$ and/or smaller
$\fgamma$.

\section{Conclusions}
\label{sec:conc}

We have made use of hydrodynamical simulations of the IGM to create
model quasar absorption spectra, and compared these with observed Keck
ESI spectra of three $z>6.2$ quasars: \qnamefourtwo\ ($z=6.42$),
\qnametwoeight\ (z=6.28), and \qnametwotwo\ (z=6.22). We compared the
CPDFs of observed Ly$\alpha$ optical depths to those generated from the
simulation by assuming various values for the size of the quasar's
surrounding HII region, $R_S$ (in comoving Mpc), volume-weighted
neutral hydrogen fraction in the ambient IGM, $\volavenf$, and the
quasar's emission rate of ionizing photons, $\dot N_Q$ (in 10$^{57}$
s$^{-1}$).  This approach has not yet been applied to the spectra of the $z\gsim 6$ quasars, and we have emphasized some important caveats, most importantly our neglect of radiative transfer effects, and our assumption that the intrinsic emission line of the quasars is symmetric around the line center.
Since large sightline-to-sightline fluctuations in IGM
properties are likely at these high-redshifts, we analyzed each quasar
independently.  We found that our approach can constrain $R_S$ to
within $\sim$ 10\%, and $\dot N_Q$ to within a factor of $\sim$ 2 for
all three quasars.  Fairly strong constraints on $\volavenf$ are
obtained from \qnametwoeight\ and \qnametwotwo, while $\volavenf$ is
unconstrained by \qnamefourtwo.  Specifically, our results are as
follows.

{\it \qnamefourtwo}: the parameter combination with the maximum
likelihood is ($R_S$, $\volavenf$, $\dot N_Q$) = (40, 0.16, 2.1); the
range of parameter values whose likelihood estimates are within a
factor of three of the peak is encompassed by: $40 \lsim R_S \lsim
42$, $\volavenf \lsim 1$, $0.9 \lsim \dot N_Q \lsim 2.5$.

{\it \qnametwoeight}: the parameter combination with the maximum
likelihood is ($R_S$, $\volavenf$, $\dot N_Q$) = (41, 1, 1.3); the
range of parameter values within a factor of three of the peak
likelihood is encompassed by: $41 \lsim R_S \lsim 45$, $0.033 \lsim
\volavenf$, $0.9 \lsim \dot N_Q \lsim 1.7$.

{\it \qnametwotwo}: the parameter combination with the maximum
likelihood is ($R_S$, $\volavenf$, $\dot N_Q$) = (29, 1, 0.9); the
range of parameter values within a factor of three of the peak
likelihood is encompassed by: $25 \lsim R_S \lsim 29$, $0.033 \lsim
\volavenf$, $0.5 \lsim \dot N_Q \lsim 1.7$.

It is encouraging that the simple methodology we have utilized can
deliver statistically interesting constraints, even from the few dozen
available spectral pixels. Our method is different from previous
analyses of the GP absorption spectra of these quasars, and the lower
limits we found on the neutral fraction of the IGM surrounding two of
the quasars strengthen the evidence that the rapid end--stage of
reionization is occurring near $z\sim 6$. Clearly, it will be
interesting to apply a similar analysis to a larger sample of sources
in the future.

\acknowledgments{The authors thank Renyue Cen for permitting the use
  of his simulation, and Xiaohui Fan for providing the quasar spectra
  used in this paper as well as for numerous helpful conversations.  The authors also thank Steven Furlanetto and Arlin Crotts for useful discussions.
  AM acknowledges support by NASA through the GSRP grant NNG05GO97H.
  ZH acknowledges partial support by NASA through grants NNG04GI88G
  and NNG05GF14G, by the NSF through grants AST-0307291 and
  AST-0307200, and by the Hungarian Ministry of Education through a
  Gy\"orgy B\'ek\'esy Fellowship.
}

\bibliographystyle{apj}
\bibliography{apj-jour,ms}

\end{document}